\documentclass[twocolumn]{aastex631}

\graphicspath{{./}{figures/}}

\usepackage{lineno}
\usepackage{csquotes}
\usepackage{gensymb}
\usepackage{chngcntr}
\usepackage{url}

\received{2023 April 27}
\revised{2023 May 16}
\accepted{2023 May 17}
\submitjournal{ApJ}

\graphicspath{{./}{figures/}}

\begin{document}

\title{The JCMT BISTRO Survey: Studying the Complex Magnetic Field of L43}


\author[0000-0001-5996-3600]{Janik Karoly}
\affiliation{Jeremiah Horrocks Institute, University of Central Lancashire, Preston PR1 2HE, UK}

\author[0000-0003-1140-2761]{Derek Ward-Thompson}
\affiliation{Jeremiah Horrocks Institute, University of Central Lancashire, Preston PR1 2HE, UK}

\author[0000-0002-8557-3582]{Kate Pattle}
\affiliation{Department of Physics and Astronomy, University College London, WC1E 6BT London, UK}

\author[0000-0001-6524-2447]{David Berry}
\affiliation{East Asian Observatory, 660 N. A{'}oh\={o}k\={u} Place, University Park, Hilo, HI 96720, USA}

\author[0000-0002-1178-5486]{Anthony Whitworth}
\affiliation{School of Physics and Astronomy, Cardiff University, The Parade, Cardiff, CF24 3AA, UK}

\author[0000-0002-4552-7477]{Jason Kirk}
\affiliation{Jeremiah Horrocks Institute, University of Central Lancashire, Preston PR1 2HE, UK}

\author[0000-0002-0794-3859]{Pierre Bastien}
\affiliation{Centre de recherche en astrophysique du Qu\'{e}bec \& d\'{e}partement de physique, Universit\'{e} de Montr\'{e}al, C.P. 6128 Succ. Centre-ville, Montr\'{e}al, QC, H3C 3J7, Canada}

\author[0000-0001-8516-2532]{Tao-Chung Ching}
\affiliation{National Radio Astronomy Observatory, 1003 Lopezville Road, Socorro, NM 87801, USA}

\author[0000-0002-0859-0805]{Simon Coud\'{e}}
\affiliation{Department of Earth, Environment, and Physics, Worcester State University, Worcester, MA 01602, USA}
\affiliation{Center for Astrophysics $\vert$ Harvard \& Smithsonian, 60 Garden Street, Cambridge, MA 02138, USA}

\author[0000-0001-7866-2686]{Jihye Hwang}
\affiliation{Korea Astronomy and Space Science Institute, 776 Daedeokdae-ro, Yuseong-gu, Daejeon 34055, Republic of Korea}
\affiliation{University of Science and Technology, Korea, 217 Gajeong-ro, Yuseong-gu, Daejeon 34113, Republic of Korea}

% ...

\begin{abstract}
We present observations of polarized dust emission at 850\,$\mu$m from the L43 molecular cloud which sits in the Ophiuchus cloud complex. The data were taken using SCUBA-2/POL-2 on the James Clerk Maxwell Telescope as a part of the BISTRO large program. L43 is a dense ($N_{\rm H_2}\sim 10^{22}$--10$^{23}$\,cm$^{-2}$) complex molecular cloud with a submillimetre-bright starless core and two protostellar sources. There appears to be an evolutionary gradient along the isolated filament that L43 is embedded within, with the most evolved source closest to the Sco OB2 association. One of the protostars drives a CO outflow that has created a cavity to the southeast. We see a magnetic field that appears to be aligned with the cavity walls of the outflow, suggesting interaction with the outflow. We also find a magnetic field strength of up to $\sim$160$\pm$30\,$\mu$G in the main starless core and up to $\sim$90$\pm$40\,$\mu$G in the more diffuse, extended region. These field strengths give magnetically super- and sub-critical values respectively and both are found to be roughly trans-Alfv\'enic. We also present a new method of data reduction for these denser but fainter objects like starless cores.
\end{abstract}


\section{Introduction} 
\label{sec:intro}
Magnetic fields (B-fields) are known to be prevalent throughout the interstellar medium (ISM) and thread through molecular clouds \citep{2016A&A...586A.138P}. Multiple simulations have demonstrated that turbulence and magnetic fields often play a role in the formation of filaments and molecular clouds \citep{fed2015}, and although the magnetic field does not appear to dominate as heavily over gravity or turbulence as first thought, it has a non-negligible influence \citep{2019FrASS...6....5H,krumholz2019}. 

Most observations of magnetic fields to date have been carried out in nearby, large star-forming regions that may already contain stars or are bright and massive \citep[e.g.][]{2017ApJ...846..122P,Soam_2018,2021A&A...647A..78A,2022ApJ...926..163K}. Early observations of dim, prestellar cores were made by \citet{2000ApJ...537L.135W}, but in more recent cases, due to increased sensitivity of polarimeters such as POL-2 on the James Clerk Maxwell Telescope (JCMT), more complex magnetic fields have been observed in molecular clouds where stars have yet to be formed \citep{2019ApJ...877...43L,2020ApJ...900..181K, 2021ApJ...907...88P}. It is important to understand the strength and structure of the magnetic fields in these early stages of star formation since their role may differ from roles played once stars have already formed, for example due to interaction with stellar feedback \citep{krumholz2019}. Additionally, magnetic fields most likely play a role in forming filamentary structures \citep{Soler_2013} which can then fragment into star-forming regions or individual stars. Bfields In STar Forming Regions Observations (BISTRO; \citealt{Ward_Thompson_2017}) is a large program on the JCMT that uses POL-2 to observe the magnetic field in star-forming regions at 850 and 450\,$\mu$m in order to understand these roles. 

Magnetic fields, however, are subject to the influence of numerous processes in molecular clouds. Gravity and turbulence are two factors which can affect the magnetic field structure and strength \citep{2019FrASS...6....5H}. Additionally, protostellar outflows have been known to either affect, or be affected by, magnetic fields as seen by many instances of the magnetic fields tracing outflows \citep[see][]{2017ApJ...847...92H,Hull_2020,2021ApJ...918...85L,2022MNRAS.515.1026P}.

In this work, we investigate the contribution of the magnetic field to the stability of the starless core within L43. We also investigate the interaction of the magnetic field with the CO outflow of RNO 91, an embedded Class I protostar. This is achieved by using 850\,$\mu$m polarization observations obtained with the POL-2 polarimeter at the JCMT. The morphology of the plane-of-sky (POS) component of the mean magnetic field (averaged along the line of sight) in the interstellar medium can be directly inferred from the polarization of dust thermal emission at far-infrared and sub-millimetre wavelengths \citep[see][and references therein]{ALV2015}. Such polarized emission is expected to be perpendicular to the plane-of-sky magnetic field orientation due to the alignment of interstellar dust grains with magnetic fields through Radiative Alignment Torques (RATs) \citep{2007MNRAS.378..910L,ALV2015} and (sub-)mm polarization parallel to the grains' main axis in the Rayleigh regime \citep{2019MNRAS.488.1211K}.

This paper is structured as follows: Section\,\ref{sec:l43} is an introduction to the L43 molecular cloud, Section\,\ref{sec:obsanddata} presents the observations and the data reduction process. Section\,\ref{sec:result} provides a discussion of the main results, such as the dust and outflow properties (\S\ref{subsec:850dust},  \ref{subsec:outflow} and \ref{subsec:cd_temp}), the polarization properties (\S\ref{subsec:pol_prop}) and the magnetic field morphology and strength (\S\ref{subsec:mag_structure} and \ref{subsec:mag_strength}). Section\,\ref{sec:disc} discusses the relation of magnetic fields with other properties in L43 such as gravity and kinematics (\S\ref{subsec:bfield_str}), the alignment of the magnetic field with the outflow (\S\ref{subsec:mag_co}) and how the filament may have evolved over time (\S\ref{subsec:filevo}). Finally, we summarize the findings of this paper in Section\,\ref{sec:summary}.

\section{Lynds 43} \label{sec:l43}

\begin{figure}
    \centering
    \includegraphics[width=0.5\textwidth]{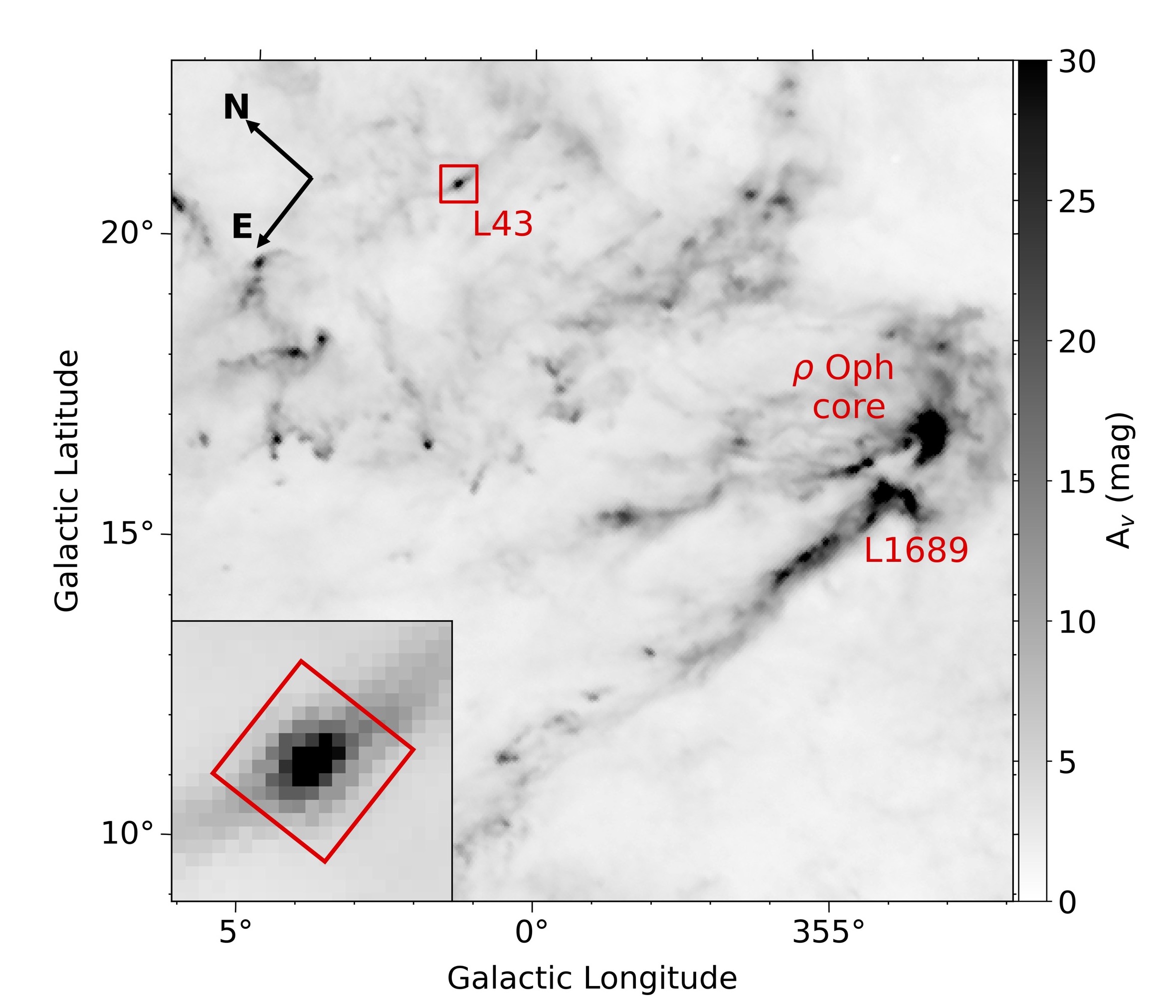}
    \caption{An extinction map of the Ophiuchus region made from Planck dust emission maps \citep{2016A&A...586A.132P}. The inset is a zoomed in picture of the red box labeled L43. The rotated red box in the inset shows the region plotted in Figure~\ref{fig:herschel} in the J2000 coordinate system. The well known clouds of the $\rho$ Oph core (also known as L1688) and L1689 are labelled as a reference.}
    \label{fig:av}
\end{figure}

\begin{figure}
    \centering
    \includegraphics[width=0.50\textwidth]{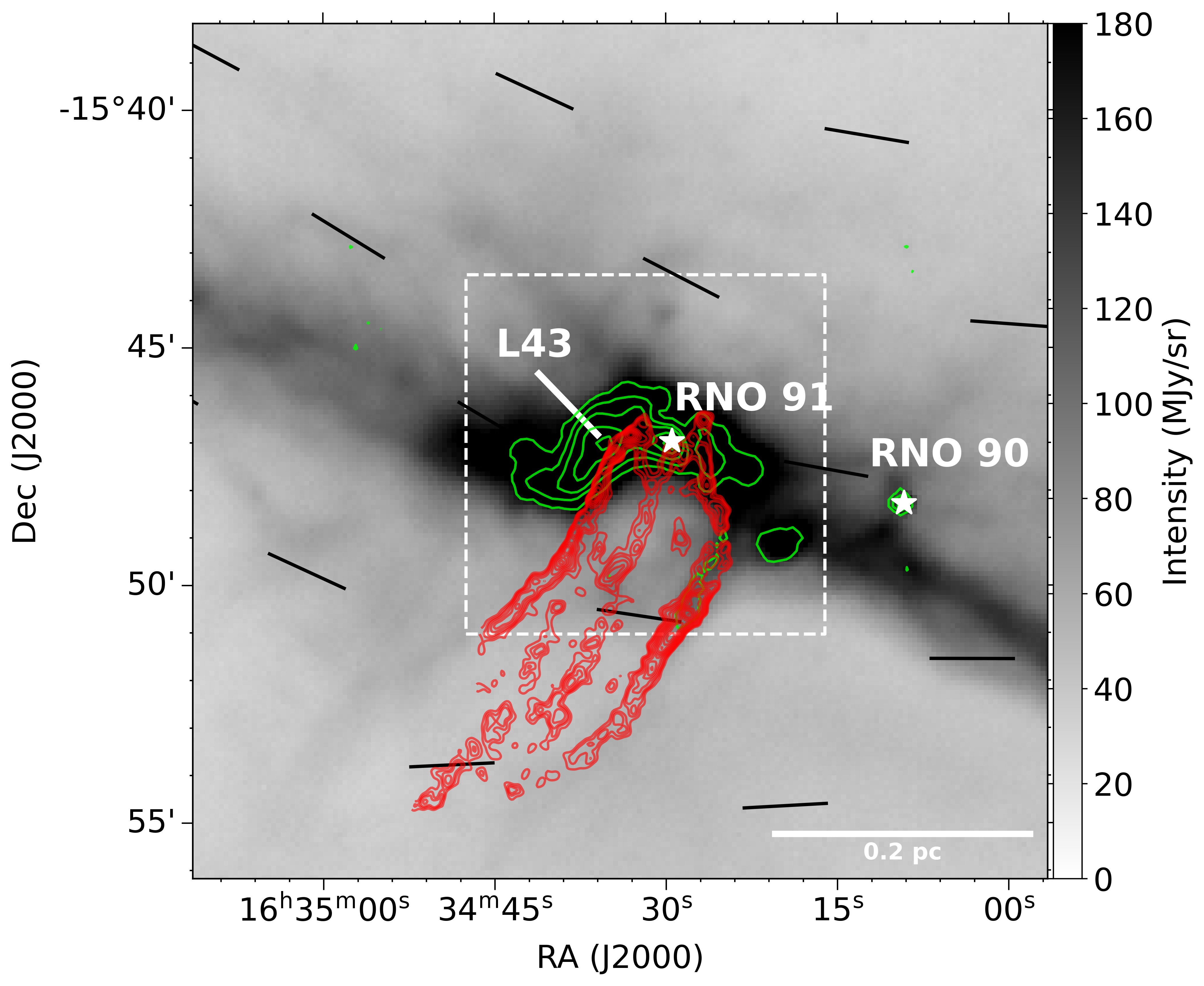}
    \caption{\textit{Herschel} SPIRE 250\,$\mu$m dust continuum map with SCUBA-2/POL-2 850\,$\mu$m dust continuum green contours from this work. Planck B-field vectors are overlaid in black and are all normalized to a single length and over-sampled at every 5$\arcmin$. The two embedded YSOs are labelled. Additionally, the CO J=1-0 emission from RNO 91 \citep{2002ApJ...576..294L} is shown in red. The white dashed box shows the area of interest that is plotted in later figures}
    \label{fig:herschel}
\end{figure}

L43 is a nearby molecular cloud in the northern region of the Ophiuchus star-forming region (see Figure\,\ref{fig:av}) at 120--125\,pc which is the mean distance to the Ophiuchus complex \citep{1990A&A...231..137D,2008ApJ...675L..29L}. As can be seen in Figure\,\ref{fig:av}, L43 is an isolated dense core with a visual extinction $>$30\,mags. It contains a sub-millimetre bright starless core \citep{2000ApJ...537L.135W} to the east and to the west an embedded young stellar object (YSO), IRAS 16316-1450, a T Tauri star \citep{1981AJ.....86..885H} originally classified as a Class II source currently transitioning from a protostar to a main-sequence star \citep{1994ApJ...420..837A}. However, \citet{Chen_2009} and \citet{2021ApJ...919..116Y} have more recently classified it as a Class I source based on Spitzer and spectral line data respectively. IRAS 16316-1450 is most commonly known as red nebulous object (RNO) 91 \citep{1980AJ.....85...29C}, although this technically refers to the reflection nebula with which the YSO is associated \citep{1994ApJS...94..615H}. The YSO is also associated with an extended, asymmetrical and bipolar CO outflow \citep[][and see Figure\,\ref{fig:herschel}]{2002ApJ...576..294L} and HCO$^{+}$, N$_2$H$^{+}$ and CS emission \citep{2005ApJ...624..841L}. The CO outflow is detected in the $^{12}$CO J = 1-0, 2-1 and 3-2 transitions, but there is no detection in the higher transitions \citep{2018ApJ...860..174Y}. The CO J = 1-0 outflow is shown in Figure \ref{fig:herschel}. The J = 2-1 transition is plotted in Figure 1 of \citet{1998mnras...299..965}. All three of the transitions show a very dominant southern outflow, although the HARP CO J = 3-2 data shows a smaller northern lobe as well.

Another YSO, similarly named RNO 90 \citep[][also known as V1003 Oph]{1980AJ.....85...29C} sits further to the west, $\sim$0.2\,pc away from RNO 91 (see Figure\,\ref{fig:herschel}), and is also classified as a T Tauri star \citep{1981AJ.....86..885H} but is a much more evolved source, with an age of 2-6\,Myr \citep{garufi22} and a protoplanetary disk \citep[e.g.][]{pont2010}. It sits at a distance of 114.7--116.7\,pc \citep{gaia21,bailerjones18} suggesting that this star sits either in the foreground of the L43 molecular cloud or perhaps the filament where they are embedded is inclined towards us (assuming the distance to L43 is similar to the mean distance of the larger Ophiuchus region, 120--125\,pc). The presence of a reflection nebula for both RNO 90 and 91 suggests they do sit just in front of or are partially embedded in the filament/molecular cloud \citep{1981AJ.....86..885H}. L43 is therefore a unique environment which consists of an older T-Tauri star, a younger Class I protostar and a starless core within a very isolated filament and molecular cloud, and with an evolutionary gradient from southwest to northeast.

Figure \ref{fig:herschel} shows the dense starless core with green contours as observed by JCMT at 850\,$\mu$m, which is embedded within a longer more diffuse filament seen by \textit{Herschel}. This isolated filament, seen also in Figure~\ref{fig:av} is oriented at $\approx$ 67$^{\circ}$ E of N. \textit{Planck} polarization observations also show a large-scale magnetic field roughly parallel to the filament, although curving slightly to the south. The magnetic field of the starless core was previously observed using the predecessor to POL-2, SCUPOL, by \citet{2000ApJ...537L.135W} and a magnetic field strength in the core was calculated to be $\approx$160\,$\mu$G \citep{2004ApJ...600..279C}. Additionally, \citet{2000ApJ...537L.135W} suggested that the magnetic field might be affected by the outflow of the RNO 91 source, although the entire molecular cloud was not observed and RNO 91 was on the very edge of the SCUPOL observations. The southern, blue-shifted lobe of the CO outflow from RNO 91 \citep{2002ApJ...576..294L} is seen in Figure\,\ref{fig:herschel} where RNO91 and 90 are also both labelled. 

\section{Observations and Data Reductions} \label{sec:obsanddata}

\subsection{SCUBA-2/POL-2 Observations}
\label{subsec:obs}
We observed L43 at 450 and 850\,$\mu$m using the Submillimetre Common-User Bolometer Array 2 (SCUBA-2)/POL-2 on JCMT, however this work focuses only on the 850\,$\mu$m observations and we leave the 450~$\mu$m data for a potential future multi-wavelength study. The observations were taken between 2020 February and 2021 March as part of the BISTRO large survey program (Project ID: M20AL018) in its third generation of observations, `BISTRO-3'. The observations consisted of 27 repeats of $\sim$31 minutes with one observation listed as questionable which we omitted from the data reduction. The data were observed in Band 1 weather conditions with the atmospheric opacity at 225\,GHz ($\tau_{225}$) less than 0.05. The JCMT has a primary dish diameter of 15\,m and a beam size of 14$\farcs$6 at 850\,$\mu$m when approximated with a two-component Gaussian \citep{2013MNRAS.430.2534D}. The observations were performed using a modified SCUBA-2 DAISY mode \citep{2013MNRAS.430.2513H} optimized for POL-2 \citep{2016SPIE.9914E..03F} which produces a central 3$\arcmin$ region with uniform coverage with noise and exposure time increasing and decreasing respectively to the edge of the map. The 3$\arcmin$ region covers most of the L43 molecular cloud including the starless core and RNO 91. This mode has a scan speed of 8$\arcsec$ s$^{-1}$ with a half-wave plate rotation frequency of 2 Hz \citep{2016SPIE.9914E..03F}.

\subsection{Data Reduction}
\label{subsec:data}

To reduce the data, we used the Submillimetre User Reduction Facility (SMURF) package \citep{2013MNRAS.430.2545C} from the Starlink software \citep{2014ASPC..485..391C}. The SMURF package contains the data reduction routine for SCUBA-2/POL-2 observations named {\it{pol2map}}\footnote{\url{ http://starlink.eao.hawaii.edu/docs/sun258.htx/sun258ss73.html}
\url{http://starlink.eao.hawaii.edu/docs/sc22.htx/sc22.html}}.

In the first step, the raw bolometer time-streams are separated into Stokes \textit{I}, \textit{Q} and \textit{U} time-streams. The command {\it{makemap}} \citep{2013MNRAS.430.2545C} is then called to create an initial Stokes \textit{I} map from the Stokes \textit{I} time-streams. The Stokes \textit{I} map from the first step is used to create an {\tt\string AST} and {\tt\string PCA} mask at a fixed signal-to-noise (S/N) ratio which are then used to mask out non-astronomical signal (we refer to these as the auto-generated masks). The {\tt\string AST} mask is used to define background regions that are forced to zero after each iteration in order to prevent growth of spurious structures in the map. The {\tt\string PCA} mask is used to define regions that are excluded from the Principle Component Analysis which removes correlated large-scale noise components from the bolometer time-streams. These excluded regions are generally source regions since these contain uncorrelated data and would disrupt the PCA process. The second step of the reduction creates the final Stokes \textit{I}, \textit{Q} and \textit{U} maps and a polarization half-vector catalog. The polarization vectors are often referred to as `half-vectors' due to the 180$^{\circ}$ ambiguity since direction is not known (e.g. 45$^{\circ}$ is treated the same as 225$^{\circ}$). We included the parameter \textit{skyloop} and followed the same data reduction technique as \citet{2021ApJ...907...88P}. We also set the parameter {\it{pixsize}} to a value of 8 to reduce the data with 8$\arcsec$ pixels as explained in Section\,\ref{subsec:pix8}. 

We corrected for instrumental polarization (IP) in the Stokes \textit{Q} and \textit{U} maps based on the final Stokes \textit{I} map and the ``August 2019" IP polarization model\footnote{\url{ https://www.eaobservatory.org/jcmt/2019/08/new-ip-models-for-pol2-data/}}. The 850-$\mu$m Stokes \textit{I}, \textit{Q} and \textit{U} maps were also multiplied by a Flux Conversion Factor (FCF) of 748 Jy\,beam$^{-1}$\,pW$^{-1}$ to convert from pW to Jy\,beam$^{-1}$ and account for loss of flux from POL-2 inserted into the telescope. This value was calculated using the post-2018 June 30 SCUBA-2 FCF of 495 Jy\,beam$^{-1}$\,pW$^{-1}$ \citep{2021AJ....162..191M} multiplied by a factor of 1.35 to account for the additional losses in POL-2 \citep{2016SPIE.9914E..03F} and then multiplied by a factor of 1.12 to account for the 8$\arcsec$ pixels. This extra factor was determined from SCUBA-2 calibration plots\footnote{ \url{https://www.eaobservatory.org/jcmt/instrumentation/continuum/scuba-2/calibration/}}. The final Stokes \textit{I} map has an RMS noise of $\approx$2.2 mJy\,beam$^{-1}$.
To further increase the S/N of our polarization half-vectors and attempt to account for the JCMT beam size, we binned them to a resolution of 12$\arcsec$. The polarization half-vectors are also debiased as described in \citet{1974ApJ...194..249W} to remove statistical bias in regions of low S/N (see Equation \ref{eq:pol}).

The values for the debiased polarization fraction $P$  were calculated from 

\begin{equation}
\label{eq:pol}
P=\frac{1}{I}\sqrt{Q^{2}+U^{2}-\frac{Q^{2}\delta Q^{2}+U^{2}\delta U^{2}}{Q^{2}+U^{2}}}   \,\, ,
\end{equation}
\noindent
where \textit{I}, \textit{Q}, and \textit{U} are the Stokes parameters, and $\delta Q$, and $\delta U$ are the uncertainties for Stokes \textit{Q} and \textit{U}. The uncertainty $\delta P$ of the polarization degree was obtained using

\begin{equation}
\delta P = \sqrt{\frac{(Q^2\delta Q^2 + U^2\delta U^2)}{I^2(Q^2+U^2)} + \frac{\delta I^2(Q^2+U^2)}{I^4}}  \,\, ,
\end{equation}
with $\delta I$ being the uncertainty for the Stokes\,\textit{I} total intensity. 

The polarization position angles $\theta$, measured from North to East in the sky projection (North is 0$\degree$), were measured using the relation 

\begin{equation}
{\theta = \frac{1}{2}tan^{-1}\frac{U}{Q}} \, .
\end{equation}

The corresponding uncertainties in $\theta$ were calculated using

\begin{equation}
\delta\theta = \frac{1}{2}\frac{\sqrt{Q^2\delta U^2+ U^2\delta Q^2}}{(Q^2+U^2)} \times\frac{180\degree}{\pi}  \,\, .
\label{eq:dtheta}
\end{equation}

The plane-of-sky orientation of the magnetic field is inferred by rotating the polarization angles by 90$\degree$ (assuming that the polarization is caused by elongated dust grains aligned perpendicular to the magnetic field).

\subsection{Use of 8 Arcsecond Pixels}
\label{subsec:pix8}

Standard reductions of SCUBA-2/POL-2 observations are done with a 4$\arcsec$ pixel size \citep[e.g.][]{2021ApJ...907...88P}. Nearly all of these reductions have been done on bright, high S/N sources. As we approach the limit of the POL-2 polarimeter, we can explore the use of larger pixel sizes to attempt to boost the S/N in these dense, starless sources. The use of different pixel sizes in reductions of JCMT SCUBA-2 data has been explored before, such as in the Gould Belt Survey \citep{2007PASP..119..855W} where originally 6$\arcsec$ \citep[see][]{2013ApJ...767..126S} and 3$\arcsec$ \citep[see][]{2015MNRAS.454.2557M} pixels were used with the latter being chosen in order to recover small scale structure. The current pixel size of 4$\arcsec$ was picked in order to properly sample the Gaussian beam and allow the mapmaking algorithm to converge in a reasonable time \citep{2013MNRAS.430.2545C}, as well as avoid smoothing of larger pixels. However, with faint sources such as starless cores, we need to investigate the potential of using larger pixel sizes. 

One issue with using larger pixels is that the larger pixel size tends to produce masks that cover a larger area of the sky. Doubling the pixel size from 4$\arcsec$ to 8$\arcsec$ typically causes the number of bolometer samples falling in each pixel to increase by a factor of four, thus increasing the S/N of each pixel value by a factor near to two. Since each mask is defined by a fixed S/N cut-off, this causes a larger fraction of the map to be covered by the mask. An increase in the size of the {\tt\string AST} mask is potentially problematic, as it can encourage the growth of artificial large-scale structures within the masked areas \citep[see][]{2013MNRAS.430.2545C}. Distinguishing such artificial structures from real astronomical signal requires care. 

Our solution to this problem is to re-use the 4$\arcsec$ pixel auto-generated masks when creating externally masked maps with 8$\arcsec$ pixels, rather than using new masks based on the auto-masked 8$\arcsec$ maps. The smaller 4$\arcsec$ masks will then restrict the growth of artificial extended structures giving us more confidence in the remaining extended structure. To do this, we ran the entire reduction using the standard 4$\arcsec$ pixel size. We then regridded the {\tt\string AST} and {\tt\string PCA} masks from that reduction to 8$\arcsec$ using the command \textit{compave} from the KAPPA package \citep{2014ascl.soft03022C}. We then ran the second step of the reduction using the regridded {\tt\string AST} and {\tt\string PCA} masks to create the externally masked Stokes \textit{I}, \textit{Q} and \textit{U} maps as well as the polarization vector catalogs, using a pixel size of 8$\arcsec$. This resulted in a molecular cloud that looked similar to the original 4$\arcsec$ reduction but with better S/N and therefore more polarization half-vectors (vector catalog increased from 98 vectors to 133 vectors at the same S/N cut). This is the reduction which we present in this work.

As a further check, we performed a Jackknife Test by dividing our observations into two populations and comparing the Stokes \textit{I}, \textit{Q} and \textit{U} maps from both populations. The results from this test can be found in Appendix~\ref{app:pix8}. We saw a more significant difference between the populations when using the auto-masked 8$\arcsec$ maps. This difference occurred mainly in the areas where emission was present in the 8$\arcsec$ maps but not present in the 4$\arcsec$ maps, raising further doubt as to the validity of the new extended emission in the auto-masked 8$\arcsec$ maps. Any differences seen in 8$\arcsec$ reduction done using the regridded masks were the same as differences seen in 4$\arcsec$ reduction, just smoothed due to larger pixel sizes.

\subsection{CO Observations}
\label{subsec:co_obs}
We used archival observations of the CO J = 1-0 line carried out with the Berkely Illinois Maryland Array (BIMA) 10 antenna interferometry array. The CO J = 1-0 data were obtained from \citet{2002ApJ...576..294L} and details of the observations and data reduction can be found therein. BIMA has a similar beam size to that of JCMT at 12$\farcs$8. 

We also used archival observations of the CO J = 3-2 line carried out with the Heterodyne Array Receiver Program (HARP) to remove CO contribution from the 850\,$\mu$m Stokes \textit{I} map. This is discussed further in Section\,\ref{subsec:cd_temp}. The data were accessed from the Canadian Astronomy Data Centre database\footnote{\url{https://www.cadc-ccda.hia-iha.nrc-cnrc.gc.ca/en/jcmt/}} (Project ID: M07AU11) and were downloaded as reduced spectral cubes which were then mosaicked using the PICARD recipe MOSAIC\_JCMT\_IMAGES\footnote{\url{http://www.starlink.ac.uk/docs/sun265.htx/sun265ss15.html}}. 

\section{Results} 
\label{sec:result}

\subsection{850~$\mu$m Dust Morphology}
\label{subsec:850dust}
Figure~\ref{fig:herschel} shows the 850~$\mu$m dust contours in green overlaid on the \textit{Herschel} SPIRE 250~$\mu$m where the 850~$\mu$m dust traces the densest part of the filament. The filament does continue to the east and west but this may be more extended structure and is therefore lost by SCUBA-2/POL-2. The 850~$\mu$m dust traces the northern edge of the CO outflow cavity which is discussed in the next section. The 850~$\mu$m emission is peaked in the main starless core (L43) and then the dust region surrounding RNO 91.

Figure~\ref{fig:cdtemp} shows the column density map which is discussed later in Sec.~\ref{subsec:cd_temp} but it has the 850~$\mu$m dust contours overlaid with more levels to better show the emission structure. In the main starless core, the densest emission peaks toward the centre, but then there are two lobes that extend to the northwest and southeast. A small peak can be seen in southeast lobe in Figure~\ref{fig:cdtemp}. We do not have resolved kinematic or significant magnetic field data between these three areas (the centre part and the two lobes) so it is not possible to tell if they are fragmenting. However the 850~$\mu$m emission shows structure suggesting these could be on the way to fragmentation. 

We can model these three regions as Bonnor-Ebert spheres \citep{1955ZA.....37..217E,1956MNRAS.116..351B} and estimate their critical BE masses. We take the sound speed $c_{\rm s}$ to be $\sim$0.19\,km\,s$^{-1}$ which was calculated assuming a dust temperature of 12.1\,K \citep{2016A&A...594A..28P}. The critical BE mass can be calculated using the relation \citep[Eq 3.9,][]{1956MNRAS.116..351B},
\begin{equation}
M_{\rm BE, crit} = 3.3\frac{c_{\rm s}^{2}}{G}R_{\rm crit} \;,
\label{eq:becrit}
\end{equation}
\noindent
where $G$ is the gravitational constant and $R_{\rm crit}$ is the critical radius of the core. We estimate $R_{\rm crit}$ from the observed flux structure (where the flux drops before peaking again in the center) and use it to also calculate total flux for total mass estimates. Values for $R_{\rm crit}$ are given in Table~\ref{tab:tab4} along with locations of the three potentially fragmenting regions in the submillimeter core (NW, Main and SE). The estimated critical BE masses of the two lobes are $M_{\rm BE, crit}^{\rm NW}\sim$0.21\,M$_\odot$ and $M_{\rm BE, crit}^{\rm SE}\sim$0.20\,M$_\odot$ for the northwest and southeast lobes respectively. For the central (main) peak, the estimated critical BE mass is $M_{\rm BE, crit}^{\rm main}\sim$0.32\,M$_\odot$

We can estimate the total mass from the 850~$\mu$m dust emission using the relation from \citet{1983QJRAS..24..267H},
\begin{equation}
M_{\rm TOT} = \frac{F_\nu D^2}{\kappa_\nu B_\nu (T_\mathrm{d})} \:,
\label{eq:dustmass}
\end{equation}
\noindent
and see also \citet{2011isf..book.....W}, where
\begin{equation}
\kappa_\nu = \kappa_o \, \left( \frac{\nu}{\nu_o} \right) ^\beta \;,
\label{eq:kap}
\end{equation}
\noindent
and $F_\nu$ is the total measured flux density at the observed frequency $\nu$, $B_\nu(T_\mathrm{d})$ is the Planck function for a dust temperature $T_\mathrm{d}$ \citep[$T_\mathrm{d}$=12.1\,K][]{2016A&A...594A..28P}, and $\kappa_\nu$ is the monochromatic opacity per unit mass of dust and gas. $\kappa_\nu\sim$0.0125 cm$^{2}$\,g$^{-1}$ assuming $\kappa_o = 0.1$\,cm$^2$\,g$^{-1}$, $\nu_{o} = 10^{12}$\ Hz \citep{1990AJ.....99..924B} and $\beta$=2. We should note that $\kappa_\nu$ can have a systematic uncertainty of up to 50\% \citep{roy2014}.

Assuming a distance of 125\,pc, we estimate total masses from the 850~$\mu$m dust emission of $M_{\rm TOT}^{\rm NW}\sim$0.10\,M$_\odot$ and $M_{\rm TOT}^{\rm SE}\sim$0.12\,M$_\odot$ for the northwest and southeast lobes respectively and $M_{\rm TOT}^{\rm main}\sim$0.37\,M$_\odot$ for the main core. This suggests that if they are indeed fragmented, the central part of the starless core may be undergoing gravitational collapse (i.e. M$_{\rm TOT}^{\rm main}$/M$_{\rm BE, crit}^{\rm main}>$1) while the two smaller lobes are not, rather than a coherent collapse of the whole core. All of the masses are summarized in Table~\ref{tab:tab4}.

We also estimated the envelope mass from the 850~$\mu$m dust emission of the two T-Tauri sources RNO 90 and RNO 91 using Eq.~\ref{eq:dustmass}. We find a total estimated envelope mass for RNO 90 of $M_{\rm TOT}^{\rm RNO 90}$=0.033$\pm$0.016\,M$_\odot$ assuming a distance of 115.7$\pm$1\,pc and a radius of 14.4$\arcsec$. This is orders of magnitude greater than the dust mass of the disk as seen by ALMA which is closer to 2$\times$10$^{-5}$\,M$_\odot$ \citep{garufi22}, but we do not resolve this structure and are more likely seeing the remaining dusty envelope. For RNO 91, we estimate a total mass from the 850~$\mu$m dust emission of $M_{\rm TOT}^{\rm RNO 91}$=$\sim$0.335$\pm$0.169\,M$_\odot$ assuming a distance of 125\,pc and that the dusty envelope is an ellipse with dimensions 32.0$\times$18.0$\arcsec$ rotated 60$\degree$ East of North. This estimated total mass value is in good agreement with \citet{young2006} who found a mass of 0.3$\pm$0.1\,M$_\odot$. Assuming just a uniform sphere of radius 15.6$\arcsec$, we get an estimated total mass of 0.215$\pm$0.109\,M$_\odot$. 

\begin{center}
\begin{table*}
    \begin{center}
    \caption{Mass Estimates}
    \scriptsize
    \begin{tabular}{cccccc}\hline
     & \multicolumn{3}{c}{SMM core sources} & \multicolumn{2}{c}{Protostellar sources} \\
     &NW Lobe$^{\rm a}$  &Main$^{\rm a}$ &SE Lobe$^{\rm a}$ &RNO 90 &RNO 91$^{\rm a}$ \\ \hline
    RA (J2000) & 16:34:32.73 & 16:34:35.33 & 16:34:37.16 &16:34:09.29 &16:34:29.57 \\
    DEC (J2000) &-15:46:31.0 &-15:46:58.72 &-15:47:32.2 &-15:48:14.9 &-15:46:58.6 \\
    $R_{\rm crit}$ ($\arcsec$) & 12.8 & 19.6 & 12.0 & -- & -- \\
    $R$ ($\arcsec$) & 12.8 & 19.6 & 12.0 & 14.4 & 32.0$\times$18.0 (60$\degree$) \\
    $M_{\rm BE, crit}^{\rm b} (M_\odot)$ &0.21 &0.32 &0.20 & -- & -- \\
    $M_{\rm TOT}^{\rm c} (M_\odot)$ &0.10 &0.37 &0.12 &0.033(0.016) &0.335(0.169) \\ \hline
    \end{tabular}
    \label{tab:tab4}
    \end{center}
    \begin{center}
    $R_{\rm crit}$ is the critical radius of the object as described in Sec.~\ref{subsec:850dust} used to estimate critical BE masses. $R$ is the observed radius of the source as based on the flux distribution. For RNO 91 we have listed the semi-major and semi-minor axes of the ellipse with the position angle (E of N) in parentheses. \\
    a. Distance to source taken to be 125\,pc (see Sec.~\ref{sec:l43}) \\
    b. See Equation~\ref{eq:becrit} \\
    c. See Equation~\ref{eq:dustmass} \\
    \end{center}
	
\end{table*}
\end{center}

\subsection{Outflow of RNO 91}
\label{subsec:outflow}
As mentioned in Section~\ref{sec:l43}, there is a weak CO outflow driven by the embedded Class I protostar in RNO 91. The southern outflow traces the southern edge of the L43 starless core and forms a limb-brightened U shape \citep{2002ApJ...576..294L}, which is seen in all transitions. The southern outflow is heavily blue-shifted, indicating the outflow is tilted towards us, potentially by up to 60$\degree$ \citep{2005ApJ...624..841L}. \citet{1994ApJ...423..674W} finds that RNO 91 is not very deeply embedded in the L43 molecular cloud and rather sits nearer to us than the main submillimetre starless core. 

However, a very clear dust cavity can be seen in Figure\,\ref{fig:herschel} which the outflow traces nearly perfectly. This dust cavity sits along the filament, and it appears that the filament has been disrupted by the outflow, as material is cleared to the south and potentially pushed north to form the kink in the filament, though there is not much redshifted CO emission to the north. This morphology of the dust in the filament suggests some sort of interaction with or influence by the outflow. This does contradict the above claim that the source is not deeply embedded. The 850~$\mu$m dust emission also shows this U-shaped bend to the south, suggesting even the densest part of the filament is affected by, or was initially affected by, the outflow. \citet{1998mnras...299..965} did suggest that the outflow has been weakened over time by a UV radiation field, so the current outflow we observe may not be the original morphology or strength. 

One possibility then is that the source was previously embedded and cleared out the dust cavity we see including along the LOS so that it presently sits in the foreground of the dense filament. This was also suggested by \citet{1988ApJ...330..385M} who determined that RNO 91 was once associated with the dense molecular core, but has since blown through the dense gas with the outflow. They also suggest that the outflow energy is only coupled with a small fraction of the core mass. So the majority of the dense starless core L43 is undisturbed, though as discussed later, our observations of the magnetic field suggest that we are either only tracing affected foreground dust or that the outflow has influenced some of the dense material.

Regardless, the fact that the dust appears to be heavily influenced by the outflow suggests we must be careful in our analysis of the magnetic field which is traced by the dust. A more in-depth discussion of the interaction of the outflow with the magnetic field and potential CO emission or polarization contribution is presented in Section\,\ref{subsec:mag_co}.

\subsection{Dust Column Density}
\label{subsec:cd_temp}
We used archival \textit{Herschel} Photodetector Array and Camera Spectrometer (PACS) 160\,$\mu$m, SPIRE 250, 350 and 500\,$\mu$m dust emission maps\footnote{from \url{http://archives.esac.esa.int/hsa/whsa/}}, along with the JCMT 850\,$\mu$m dust emission map from this work to create column density maps. We filtered the \textit{Herschel} maps in order to remove the large-scale structure that SCUBA-2/POL-2 is not sensitive to. We followed the method from \citet{2013ApJ...767..126S} and \citet{2015MNRAS.450.1094P} of introducing the \textit{Herschel} maps into the Stokes \textit{I} timestream and repeating the reduction process from Section\,\ref{subsec:data}, using 4$\arcsec$ pixels. We then subtracted the original 850~$\mu$m only SCUBA-2/POL-2 Stokes \textit{I} emission from the map which included the \textit{Herschel} maps in the reduction and the resulting map was the filtered \textit{Herschel} map.

We also attempted to `correct' the 850\,$\mu$m Stokes \textit{I} maps by removing potential contamination from the $^{12}$CO (J = 3-2) line. The $^{12}$CO (J = 3-2) line which sits at 345.796 GHz is within the SCUBA-2 850\,$\mu$m bandpass filter so can contribute to total intensity continuum. We followed the method of \citet{harrietco18} using the HARP data mentioned in Section\,\ref{subsec:co_obs}. We used the regular 4$\arcsec$ Stokes \textit{I} map because this correction method is best-characterised for 4$\arcsec$ maps before and for the purposes of fitting the black-body spectrum, we do not require the increase in signal-to-noise that is helpful for our polarization vectors. The contributionto total intensity from CO was $\sim$5--10\%, getting up to $\sim$20\% directly around RNO 91. We should note that the reduction produced slight negative bowling to the north of the L43 emission, though not in a region of any emission. 

\begin{figure}
	\centering
    \includegraphics[width=0.45\textwidth]{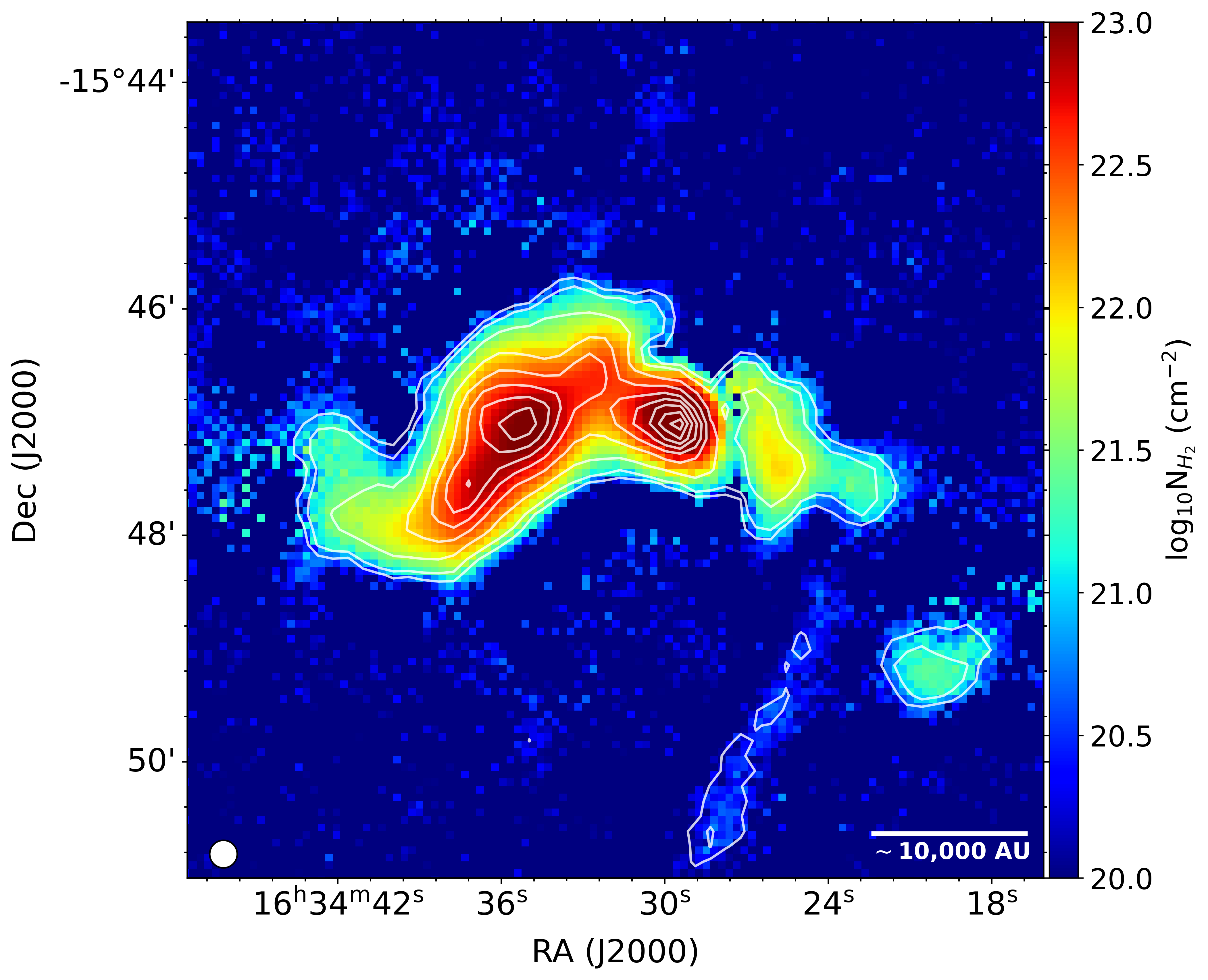}
	\caption{Molecular hydrogen column density map calculated from filtered PACS 160\,$\mu$m, SPIRE 250, 350 and 500\,$\mu$m and SCUBA-2/POL-2 850\,$\mu$m maps, with 850\,$\mu$m contours overlaid. The method for calculating the H$_{2}$ column density is described in Section~\ref{subsec:cd_temp}.}
	\label{fig:cdtemp}
\end{figure}

We then fit the five maps with a modified black-body \citep{1983QJRAS..24..267H}

\begin{equation}
F_\nu = \mu_\mathrm{H_2} m_\mathrm{H} N_\mathrm{H_2} B_\nu (T_\mathrm{d}) \kappa_\nu \:,
\label{eq:sed}
\end{equation}
\noindent
where again $F_\nu$ is the measured flux density at the observed frequency $\nu$, $B_\nu(T_\mathrm{d})$ is the Planck function for a dust temperature $T_\mathrm{d}$, $\mu_{H_2}$ is the mean molecular weight of the hydrogen gas in the cloud, $m_{H}$ is the mass of an hydrogen atom, $N_{\text{H}_2}$ is the column density, and $\kappa_\nu$ is the dust opacity (see Eq.~\ref{eq:kap}). We use a value of 2.8 for $\mu_{H_2}$, and $\kappa_\nu$ was calculated for each frequency observed using Equation \ref{eq:kap}, where $\beta$ is the emissivity spectral index of the dust and is taken to be 1.8 \citep[an approximate value in starless cores,][]{2010ApJ...708..127S,2005ApJ...632..982S,2013ApJ...767..126S}, and we again assume $\kappa_o = 0.1$\,cm$^2$\,g$^{-1}$ and $\nu_{o} = 10^{12}$\ Hz \citep{1990AJ.....99..924B}. We used temperature values from previously derived dust temperature maps using just the non-filtered SPIRE maps and 850\,$\mu$m maps, where we had convolved the data to the 500\,$\mu$m resolution of $\sim$35$\arcsec$ and then regridded to the 850\,$\mu$m maps.

We see column densities in the main starless core on the order of 10$^{22.8}$\,cm$^{-2}$ which is $\sim 6\times10^{22}$\,cm$^{-2}$, with a maximum column density of $\sim3\times10^{23}$\,cm$^{-2}$

\subsection{Polarization Properties of the Starless Core}
\label{subsec:pol_prop}
\begin{figure}
	\centering
    \includegraphics[width=0.45\textwidth]{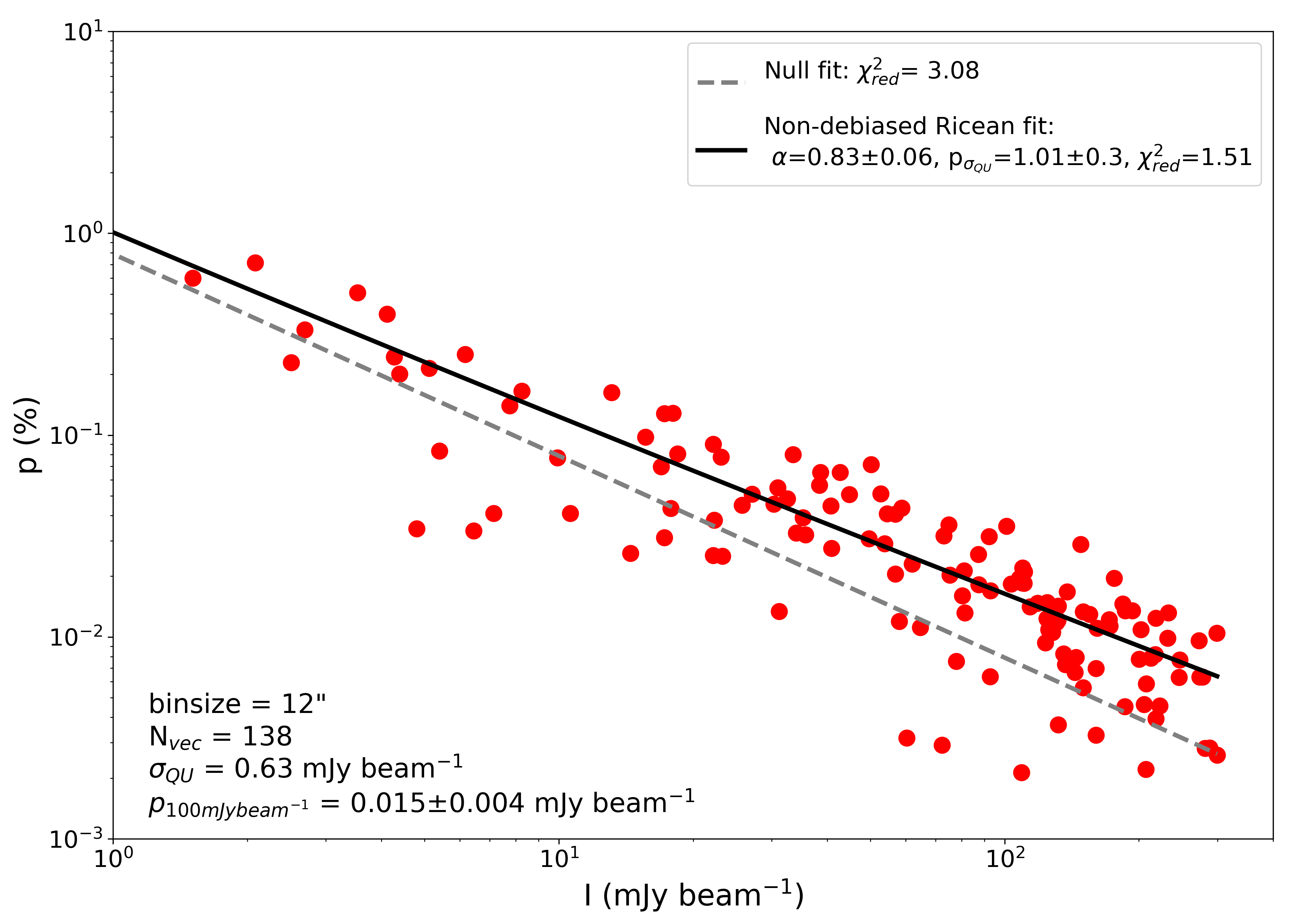}
    \caption{A plot of polarization fraction versus Stokes \textit{I} intensity of non-debiased polarization vectors in the inner 3$\arcmin$ area of the map. The vectors are binned to 12$\arcsec$ and the only selection criteria is Stokes \textit{I} $> 0$. The null fit is plotted as a grey dashed line, while the Ricean fit is plotted as a black solid line. The $\alpha$ value for the Ricean and reduced-$\chi^{2}$ values are given for both fits in the legend.}
	\label{fig:pvsav}
\end{figure}

In Figure\,\ref{fig:pvsav} we plot polarization fraction versus intensity of the non-debiased polarization half-vectors in L43. We focus only on the central 3$\arcmin$ region of L43 as this is where the exposure and noise are roughly uniform. A very clear decrease in polarization fraction can be seen towards the regions of high intensity. Within starless cores, this depolarization occurs in the highest density regions, due to some combination of field tangling and the loss of grain alignment at high enough A$_{V}$'s ($\approx$20\,mag) as predicted by RAT theory \citep{ALV2015}. A common method to study the grain alignment efficiency in molecular clouds is to determine the relationship between polarization efficiency and visual extinction, where polarization fraction and total intensity can be substituted for those two quantities respectively at submillimetre wavelengths (see \citealt{Pattle_2019} and references therein).

The relationship between polarization and intensity should follow a power law, p $\propto$ A$_{V}^{-\alpha}$, where an $\alpha$ of 1 indicates a loss of alignment and an $\alpha$ of 0 would indicate perfect alignment. We follow the methods of \citet{Pattle_2019} and use the Ricean fitting technique to fit the data. We get $\alpha$=0.83$\pm$0.06 for the 12$\arcsec$ vectors and see an obvious offset from the null which would indicate we retain some alignment. Additionally, the ordered polarization geometry suggests that we are continuing to trace the magnetic field to high A$_{V}$'s. We performed the Ricean fitting for polarization vectors binned from 8$\arcsec$ up to 32$\arcsec$ and see $\alpha$ values from 0.89 to 0.70 respectively.

\subsection{Magnetic Field Morphology}
\label{subsec:mag_structure}

\begin{figure}
    \centering
    \includegraphics[width=0.45\textwidth]{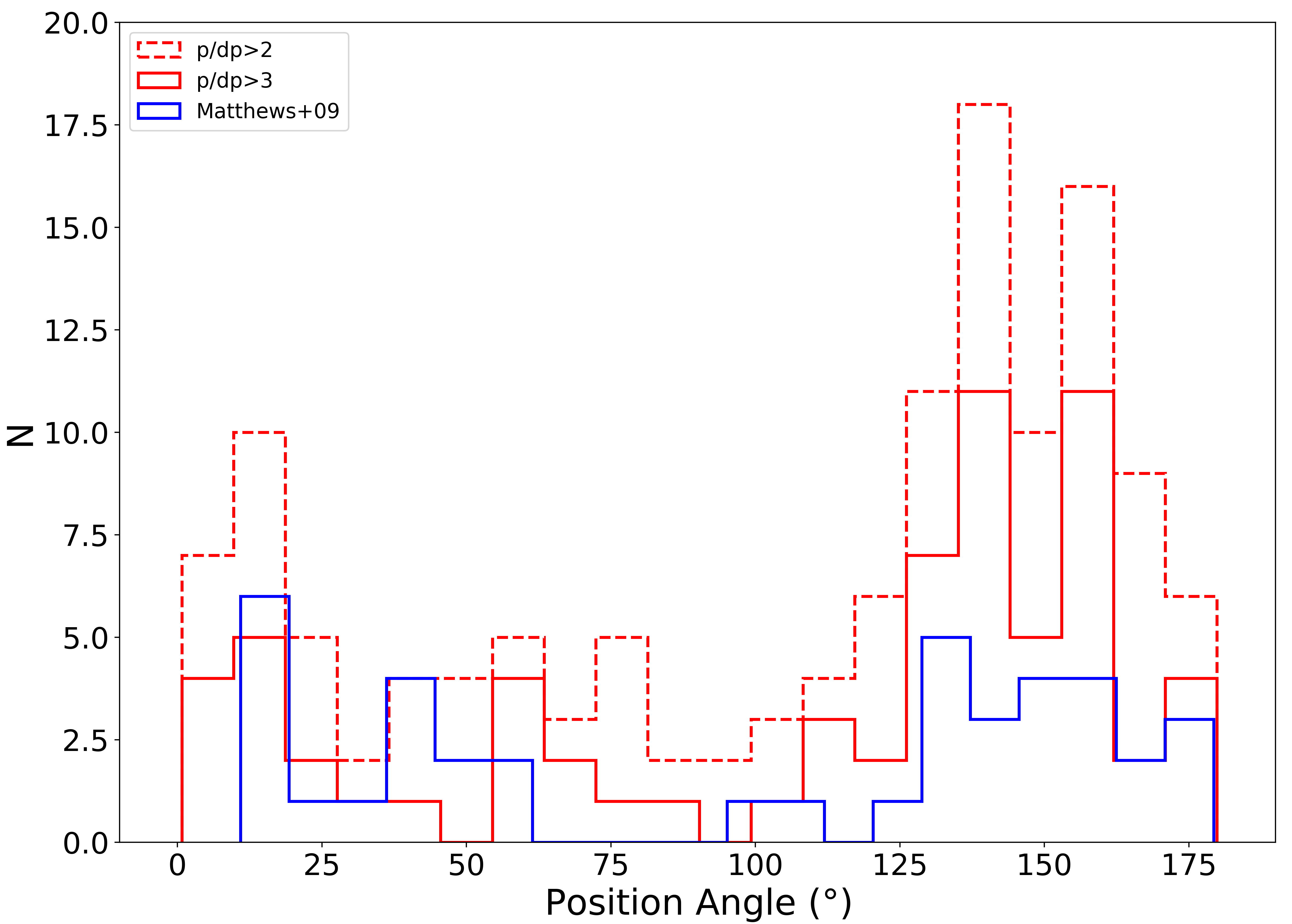}
    \caption{The distribution of the position angles of polarization half-vectors rotated by 90$\degree$ to infer the magnetic field orientation. The dashed line distribution has an S/N cut of $I/\delta_{I} > 10$ and $p/\delta_{p} > 2$. The solid red line distribution has a stricter $p/\delta_{p} > 3$ S/N cut applied. The B-field position angle distribution from \citet{matthews09} are plotted in blue. The vector populations between the two S/N cuts appear consistent suggesting that the lower S/N vectors still trace the magnetic field. They also agree with the previous SCUPOL observations \citep{matthews09}.}
    \label{fig:S/N}
\end{figure}

The vectors chosen for analysis have a S/N cut of $I/\delta_{I} > 10$ and $p/\delta_{p} > 2$. A S/N cut of $p/\delta_{p} > 2$ can be quite poor in polarization so we must proceed with caution with those vectors. In Figure~\ref{fig:S/N} we plot the lower S/N vector distribution (dashed histogram) and the higher S/N vectors with $p/\delta_{p} > 3$ (solid histogram). Within most of the molecular cloud, the two S/N cuts agree well with the lower S/N vectors following the same orientation as the higher S/N vectors. The polarization angle distributions follow the same shape between the two S/N cuts and they agree well with that found by \citet{matthews09} in the SCUPOL legacy survey (blue histogram). Using the lower S/N cut we get more data points to then use when calculating magnetic field strength (see Sec.~\ref{subsec:mag_strength}) which could increase the spread of the position angles but can also increase the statistical confidence in the calculated dispersion.

As can be seen in Figure~\ref{fig:S/N}, there is no clear single morphology of the magnetic field and it instead must be considered as either randomized or a multiple-component field. There is a rather distinct peak around 150$\degree$ with then more scatter towards the lower magnetic field polarization angles. Some of this scatter is structured fields in other parts of the molecular cloud. As will be discussed later in Section \ref{subsec:mag_co}, we suspect that the magnetic field is partially influenced by the CO outflow from RNO 91. This was discussed as well in \citet{2000ApJ...537L.135W} where they suggested the western edge of the field they observed was being influenced by RNO 91. With the more sensitive POL-2 observations and a larger FOV, we can actually see the overlap of the CO emission with some of the magnetic field vectors.

We split the magnetic field inferred from the 850\,$\mu$m polarized emission of L43 into three parts, the two labeled regions seen in Figure\,\ref{fig:b_regions} and then the magnetic field vectors which spatially (in the plane-of-sky) overlap with the CO outflow or are nearby and follow the same orientation. We list the mean field orientation, $\langle\theta_{\rm B}\rangle$, and standard deviation from a Gaussian fit of the magnetic field position angle distributions in Table~\ref{tab:tab3}. Regions 1 and 2 show different magnetic field orientations, although both are rather scattered. Region 2 which corresponds to the northern half of the starless core has a magnetic field that has an average orientation of 63$\degree$ E of N which is roughly parallel to the filament ($\approx$67$\degree$) and \textit{Planck} magnetic field orientations ($\approx$60$\degree$). It also lies roughly perpendicular to the local core elongation axis which is something seen across starless and prestellar cores \citep[see][for a recent review]{2022arXiv220311179P}. This is further discussed in the context of the region's evolution in Section~\ref{subsec:filevo}. There is more scatter towards the center of the region, which is what causes the spread we see in the position angles, but the structured component can be seen on either side. 

Region 1 has a slightly more coherent magnetic field structure that is orientated $\approx$140$^{\circ}$ E of N, nearly perpendicular to the filament direction and parallel to the CO outflow. Considering it is still near to the CO outflow and is a less dense region, the magnetic field could still be influenced by the CO outflow, or we are simply seeing another component of the complex magnetic field. The mean field orientation of the vectors spatially overlapping with the CO outflow is 146$\degree$ which is well aligned with the outflow direction which we have taken to be $\sim$150$\pm$10$\degree$ due to it curving slightly.

We also detect a few B-field vectors in the dust envelope of RNO 90 and in a very diffuse `blob', isolated to the west. RNO 90 is shown in the inset of Figure~\ref{fig:b_regions} and the magnetic field is orientated roughly north-south. The magnetic field in the diffuse blob to the west appears to still follow the large-scale \textit{Planck} field, something that has been seen in diffuse cores \citep{derekl1495} and other isolated starless cores \citep[L1689B,][]{2021ApJ...907...88P}. The fact that this more diffuse region still follows the \textit{Planck} field while Region 1 does not suggests that Region 1 may indeed be, or have been, affected by the outflow.

\subsection{Magnetic Field Strength}
\label{subsec:mag_strength}

We estimated the magnetic field strength in L43 using the Davis-Chandrasekhar-Fermi (DCF) method \citep{1951PhRv...81..890D,1953ApJ...118..116C}. The DCF method (see Eq.\,\ref{eq:b}) assumes that the geometry of the mean magnetic field is uniform in each region. It then assumes that deviations from this uniformity are Alfv\'enic such that the deviations are due to non-thermal gas motions. The Alfv\'enic Mach number of the gas is given by

\begin{equation}
    \mathcal{M}_{\rm A} = \frac{\sigma_{\rm NT}}{v_{\rm A}} = \frac{\sigma_{\theta}}{Q} \;,
    \label{eq:mach}
\end{equation}
\noindent
where the non-thermal deviations are quantified by a dispersion in magnetic field position angles, $\sigma_{\theta}$. $\sigma_{\rm NT}$ is the one-dimensional non-thermal velocity dispersion of the gas and Q is a correction factor that accounts for variations of the magnetic field on scales smaller than the beam and along the line-of-sight where 0$<Q<$1 \citep{2001ApJ...546..980O}. $v_{\rm A}$ is the Alfv\'en velocity of the magnetic field and so $\mathcal{M}_{\rm A}<$1 suggests the magnetic field is more important than turbulent motions (sub-Alfv\'enic) while $\mathcal{M}_{\rm A}>$1 means the turbulent motions are more important (super-Alfv\'enic). The Alfv\'en velocity is given by

\begin{equation}
    v_{\rm A} = \frac{B}{\sqrt{4\pi\rho}} = Q\frac{\sigma_{\rm NT}}{\sigma_{\theta}} \;,
    \label{eq:alfven}
\end{equation}
\noindent
where $B$ is the magnetic field strength and $\rho$ is the gas density. Since the dispersion in position angles, $\sigma_{\theta}$, is for plane-of-sky (POS) observations we can only calculate the plane-of-sky magnetic field strength, $B_{pos}$, which is given by

\begin{equation} 
    B_{pos} \approx Q\sqrt{4\pi\rho}\frac{\sigma_{\rm NT}}{\sigma_{\theta}} \;
    \label{eq:b}
\end{equation}

This can then be simplified to 

\begin{equation}
    B_{\rm pos}(\mu \rm G) \approx 18.6\,Q\sqrt{n(\rm H_{2})(\rm cm^{-3})}\frac{\Delta v_{\rm NT}(\rm km\, s^{-1})}{\sigma_{\theta}(\rm degree)} \;
    \label{eq:badf}
\end{equation}

 Typically $Q$ is taken to be 0.5 \citep[see][]{2001ApJ...546..980O,2004ApJ...600..279C} but we will consider a range of $Q$ values, 0.28$<Q<$0.62 from \citet{Liu2022FrASS} (see their Table 3) to obtain upper and lower limits of the B-field strength. Then $n(\rm H_{2}$) is the volume density of molecular hydrogen where $n(\rm H_{2}$)=$\rho$/$\mu_{\rm H_{2}}m_{\rm H}$ and $\mu_{\rm H_{2}}$=2.8 and $m_{\rm H}$ is the mass of hydrogen. $\Delta$ v$_{\rm NT}$ is the FWHM of the non-thermal gas velocity calculated by $\Delta$ v$_{\rm NT}$ = $\sigma_{\rm NT}\sqrt{8\rm{ln}2}$. As mentioned above, $\sigma_{\theta}$ is the dispersion of the position angles of the magnetic field vectors, which we calculated using an angular dispersion function as discussed later in this section. It should be noted that \citet{2004ApJ...600..279C} finds on average $B_{pos}/B\approx\pi/4$, but since this is a general statistical correction, we do not use this when calculating the magnetic field strength.

We can rewrite Equations~\ref{eq:mach} and \ref{eq:alfven} as

\begin{equation}
    \mathcal{M}_{\rm A} = 1.74\times10^{-2} \sqrt{2} \; \frac{\sigma_{\theta}(\rm degree)}{Q} \;,
    \label{eq:machsimp}
\end{equation}
and
\begin{equation}
    v_{\rm A}(\rm km\, s^{-1}) = 24.2 Q\; \sqrt{2} \; \frac{\Delta v_{\rm NT}(\rm km\, s^{-1})}{\sigma_{\theta}(\rm degree)} \;,
    \label{eq:alfsimp}
\end{equation}
respectively. We have also included a $\sqrt{2}$ factor in Equations~\ref{eq:machsimp} and \ref{eq:alfsimp} which is suggested by \citet{heilestrol05} to account for the velocity line width assumptions, specifically converting the 1-D line-of-sight velocity measurements to an approximate value suitable for estimating the POS magnetic field strength.

\begin{figure*}
    \centering
    \includegraphics[scale=0.6,angle=0]{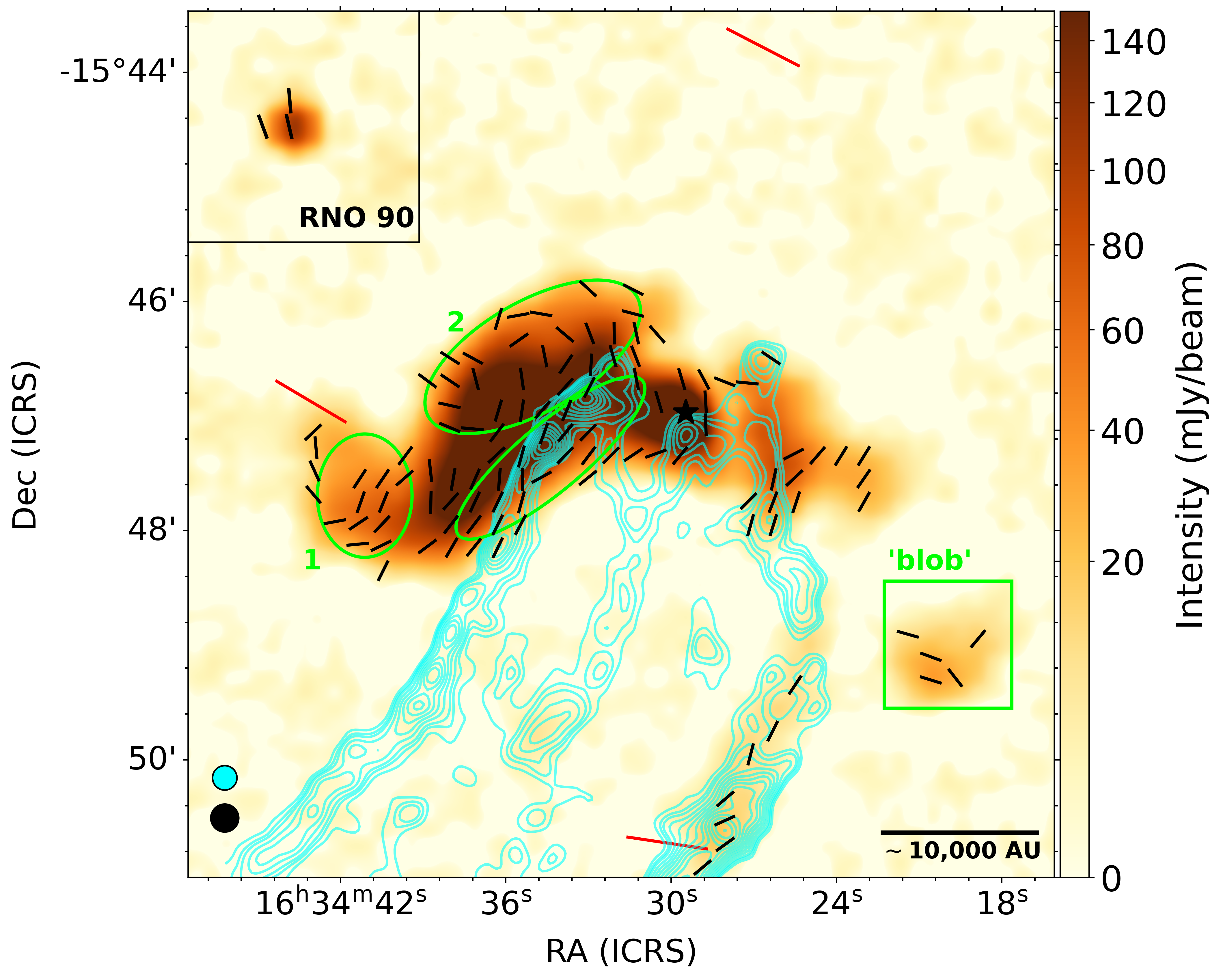}
    \caption{Magnetic field half-vectors are plotted in black with a uniform length over the 850~$\mu$m dust emission map. Planck vectors are the larger red vectors. The CO outflow continuum discussed in Sec.~\ref{subsec:outflow} is plotted with blue contours. Regions 1 and 2 are labeled and the ellipses drawn are listed in Table~\ref{tab:tab3}. The third ellipse shows the area of the cloud we used to calculate column and volume densities for the outflow vectors. We also label the dust `blob' to the west and RNO 90 is shown in the upper left corner. The BIMA and JCMT beam sizes are shown in the lower left in blue and black respectively.}
    \label{fig:b_regions}
\end{figure*}

We calculated the magnetic field strength in Regions 1 and 2 (shown in Fig.~\ref{fig:b_regions}) using Equation\,\ref{eq:badf}. We treated the regions as ellipses with semi-major and semi-minor axes a and b (see Table~\ref{tab:tab3}), and assumed the depth of those regions to be the geometric mean, $c=\sqrt{ab}$. We used column density values from Figure\,\ref{fig:cdtemp} to calculate the volume density, n(H$_{2}$), in each region. We used N$_{2}$H$^{+}$ (1-0) velocity line profiles from \citet{Caselli_2002}, which have a resolution of $\sim$0.063\,km s$^{-1}$. We corrected them to account for the thermal component (since Eq.~\ref{eq:badf} uses non-thermal velocity line widths) which was calculated using the excitation temperature, $T_{\rm ex}$=7$\pm$1\,K \citep[also from][]{Caselli_2002}, giving 0.35$\pm$0.02\,km\,s$^{-1}$. It should be noted, the velocity line profile observations are from the main starless core (Region 1). These observations do not necessarily extend to Region 2, but we do not have observations of Region 2 specifically so elect to use the same line width value as Region 1. Line widths of other tracers in the main starless core vary with some larger than and some smaller than the 0.35\,km s$^{-1}$ value we use \citep[see][]{Chen_2009}, but N$_{2}$H$^{+}$ (1-0) traces dense regions of molecular clouds which should coincide with the depths we are observing at 850\,$\mu$m as well. There are also NH$_{3}$ observations of the starless core from \citet{jijina1999} and \citet{feher2022}, with a spread of line width values from 0.273\,km\,s$^{-1}$ \citep[HFS fitting from][]{feher2022} to 0.718\,km\,s$^{-1}$ \citep[Gaussian fitting from][]{feher2022} and then 0.32\,km\,s$^{-1}$ \citep{jijina1999}. Considering the largest line width, the B-field strength could be up to two times larger.

\begin{figure}
    \includegraphics[width=0.45\textwidth]{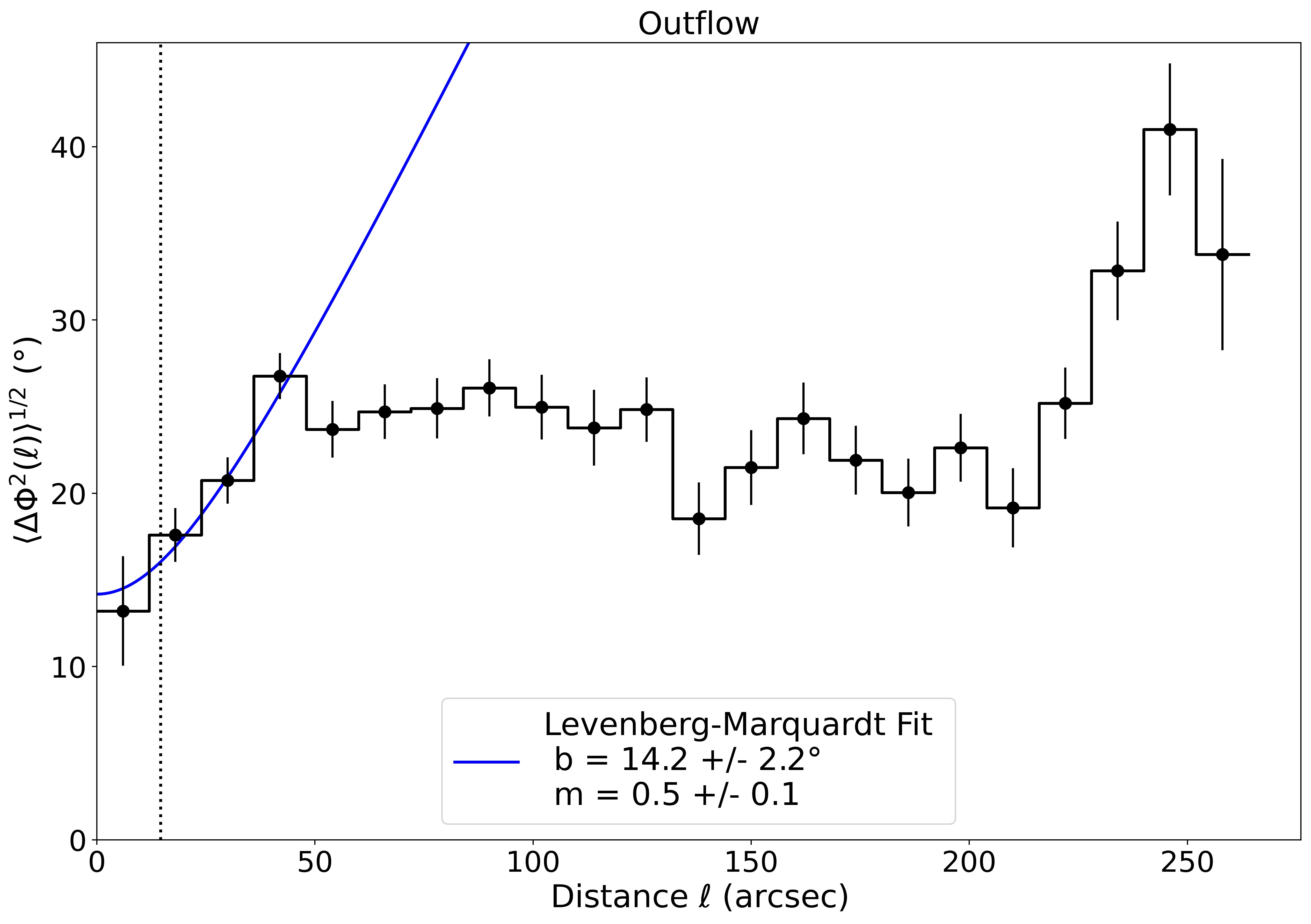}\\
    \includegraphics[width=0.45\textwidth]{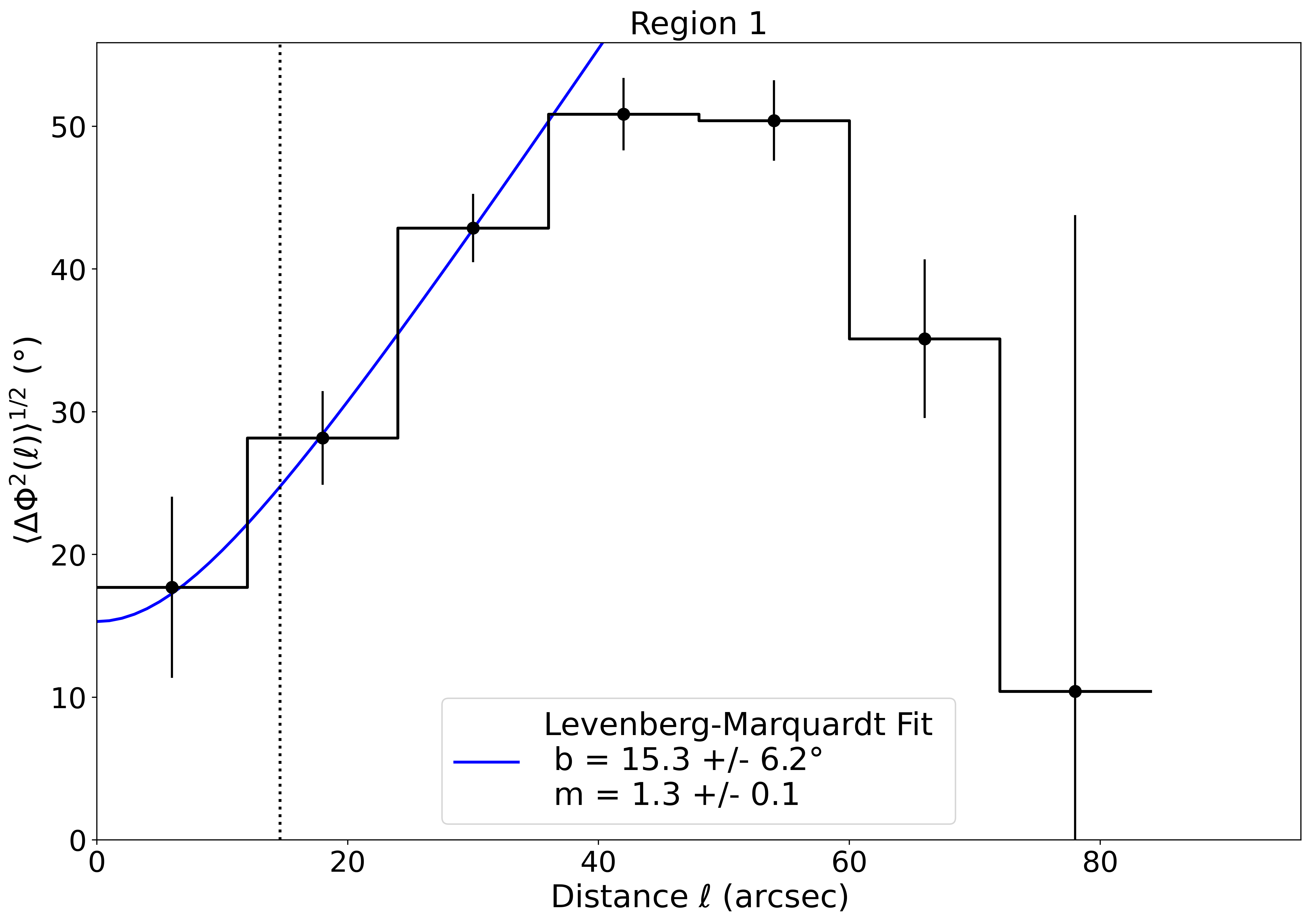}\\
    \includegraphics[width=0.45\textwidth]{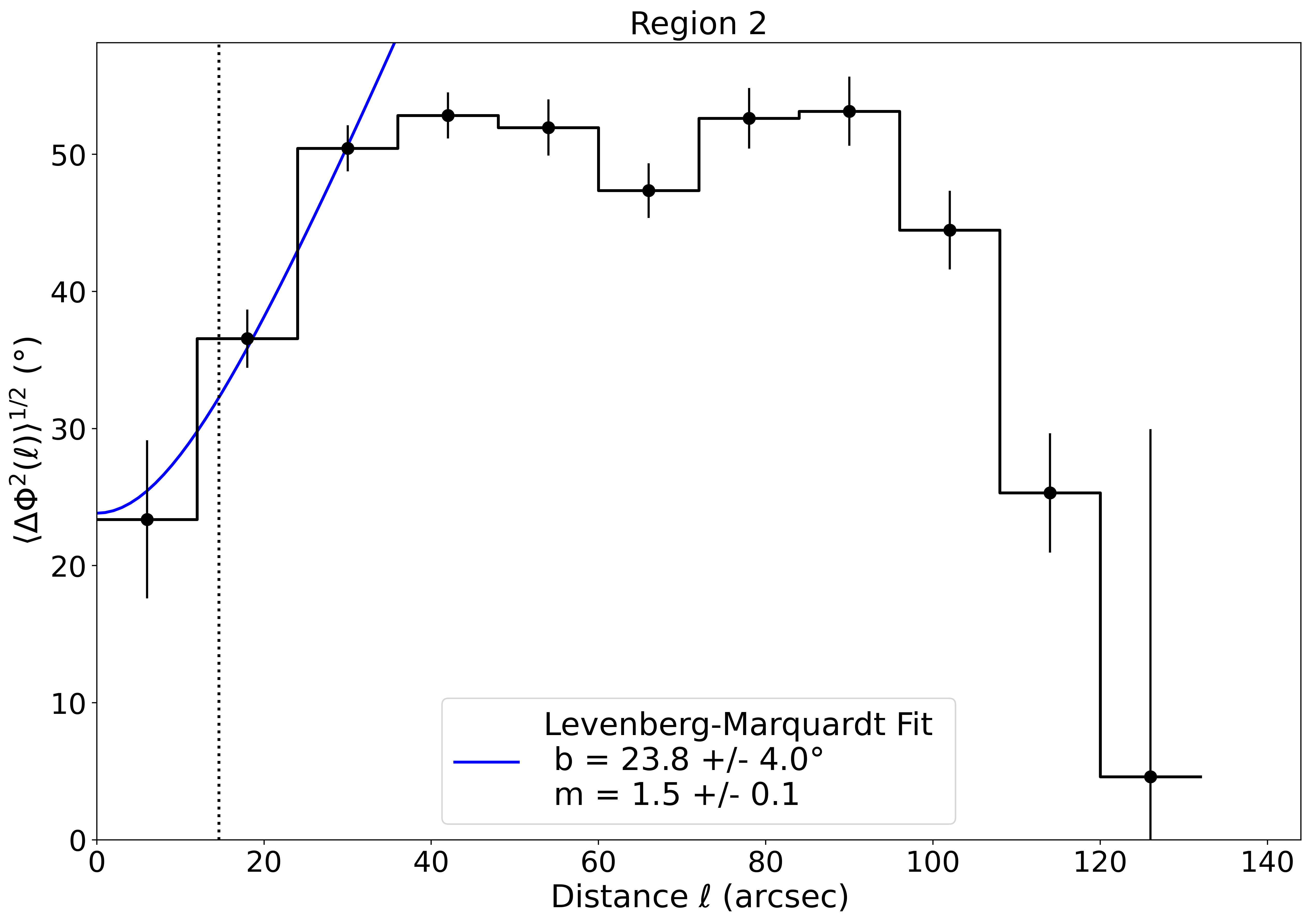}
    \caption{The angular dispersion function (ADF) histograms for various regions of interest in L43. All plots show fitting results when limiting the fit to the first three bins (36$\arcsec$). All fits were optimized with the Levenberg-Marquardt method and are weighted by the errors. Best-fit parameters are shown in the legend where $b$ and $m$ are from Eq~\ref{eq:adf}. The JCMT beam size is plotted as a vertical dotted line in all three plots. \textit{Top:} ADF results for the vectors spatially associated and aligned with the outflow. \textit{Middle:} ADF results for vectors in Region 1 (see Fig.~\ref{fig:b_regions}). \textit{Bottom:} ADF results for vectors in Region 2 (see Fig.~\ref{fig:b_regions}).}
    \label{fig:ADF}
\end{figure}

We determined the dispersion of position angles in each region using the angular dispersion function \citep[][see their Equations~1 and 3]{hild09}. This assumes that there is a large-scale structured field and a smaller turbulent or random component. We assume that the length scales we are observing with JCMT/POL-2 ($\ell$) are greater than the turbulent correlation length and much smaller than the large-scale field (such as Planck). While the former statement may be more difficult to accurately determine, the latter is true in our situation as we see a very structured large-scale B-field from Planck (see Fig.~\ref{fig:herschel}) with no variations in the region we are looking at (admittedly a single Planck beam nearly covers the whole region). The contribution of both the turbulent and large-scale components to the angular dispersion of B-field vectors $\langle\Delta\Phi^{2}(\ell)\rangle^{1/2}$ is given by the terms $b$ and $m\ell$ respectively and the relation \citep{hild09},

\begin{equation}
     \langle\Delta\Phi^{2}(\ell)\rangle_{tot} \simeq b^2 + m^{2}\ell^{2} + \delta_{\theta}^{2}(\ell) \;, 
    \label{eq:adf}
\end{equation}
\noindent
where $\delta_{\theta}^{2}(\ell)$ is the additional contribution to the dispersion from measurement uncertainties (see Eq.~\ref{eq:dtheta}). In each region we calculated $\langle\Delta\Phi^{2}(\ell)\rangle_{tot}$ and then fit Eq.~\ref{eq:adf} to determine values for $m$ and $b$. The plots are seen in Fig.~\ref{fig:ADF} where the best-fit parameters are shown as well. In all three regions, limiting the fitting to the first 3 bins (36$\arcsec$) provides the best fit (as determined by $\chi^{2} <$1), though for the outflow region, extending the fit to 48$\arcsec$ still provided $\chi^{2} <$1. For this region we then take the mean of the $b$ values from both fits (14.2 for 36$\arcsec$ and 13.7 for 48$\arcsec$) to get $b\approx$14.0$\pm$1.5. Generally we see the dispersion slowly increase with distance \citep[][and see top panel of Fig.~\ref{fig:ADF}]{hild09,hwang2023} but in the bottom two panels of Fig.~\ref{fig:ADF} we see the dispersion growing and then at a large distance, suddenly drop again. In the case of Region 2 (bottom panel of Fig.~\ref{fig:ADF}) the distance this happens at, $\sim$120$\arcsec$, is the distance between the two structured components mentioned in Sec.~\ref{subsec:mag_structure}, further justifying a structured field approximately oriented at 56$\degree$ and parallel with the Planck field and the filament direction.
The dispersion in magnetic field position angles, $\sigma_{\theta}$, can then be calculated by

\begin{equation}
    \sigma_{\theta} = \frac{b}{\sqrt{2-b^{2}}} \frac{180^o}{\pi}\;,
    \label{eq:sigtheta}
\end{equation}
\noindent
where $b$ (in radians) is found from the angular dispersion function fit \citep[see][and Fig~\ref{fig:ADF}]{hild09}.

The magnetic field strength in Regions 1 and 2 is $\sim$40--90~$\mu$G and $\sim$70--160~$\mu$G respectively. All the values calculated when considering the magnetic field strength are listed in Table~\ref{tab:tab3}. In our situation, the difference in magnetic field strength between regions is due to variations in density and angular dispersion of the vectors since we use a constant velocity line width value across the region. For example, despite the larger $\sigma_{\theta}$ in Region 2, it is also denser which is what increases the magnetic field strength and makes it comparable, if not greater than, the magnetic field in Region 1. We calculate an upper and lower bound for each of our magnetic field strengths based on the variation in $Q$ of 0.28$<Q<$0.62 \citep{Liu2022FrASS}. The value in Region 2 is approximately that found by \citet{2004ApJ...600..279C}, although they treated the region as a sphere and had a smaller dispersion angle of 12$\degree$. We find a slightly larger column and volume density, by a factor of $\sim$1.3.

The magnetic field strength calculated for the outflow is $\sim$120-260~$\mu$G. However, we note that this value may be severely overestimated and its use for interpretation limited. This is because we do not generally apply the DCF method to regions interacting with outflows since we assume the deviations in the magnetic field to be small non-thermal gas motions in the region and an outflow is a much stronger disruptive force. In our case, we are arguing that some of the dense gas and dust has been affected by the outflow and hence that the outflow may have dragged the field (which is flux-frozen into the gas), aligning it with the outflow walls and giving us the low position angle dispersion. In that case, the low position angle dispersion may be due to a strong outflow rather than a strong field; although this would require more outflow modelling to determine if the outflow would preferentially align and order the field, or disorder the field. Additionally, we consider all the vectors spatially aligned with the outflow, but take just the dust density in the southern half of the L43 starless core. If we were to assume a much lower volume density, one that is more likely associated with outflow material, then our field strength would be lower.

We acknowledge the limitations of using the DCF method, especially when determining the position angle dispersion and acknowledge in such a low S/N environment, these uncertainties are increased. However, we do find magnetic field strengths that are on the order of strengths seen in other starless cores \citep{2021ApJ...907...88P,2020ApJ...900..181K}, including values which agree within error with those found in \citet{2004ApJ...600..279C}.

\begin{center}
\begin{table*}
    \begin{center}
    \caption{DCF+ADF Values}
    \scriptsize
    \begin{tabular}{cccc}\hline
    Region &Outflow  &Reg 1 &Reg 2\\ \hline
    $\langle\theta_{\rm B}\rangle ^{a}$ ($\degree$) & 146(22) & 140(25) & 63(24) \\ 
    RA (J2000) &16:34:34.9 &16:34:41.7 &16:34:35.6 \\
    DEC (J2000) &-15:47:30.0 &-15:47:49.8 &-15:46:37.1 \\
    a ($\arcsec$) & 62.9 & 32.3 & 63.0 \\
    b ($\arcsec$) & 17.4 & 24.8 & 28.8 \\
    PA$^{b}$ ($\degree$) & 130 & 0 & 120 \\
    c ($\arcsec$) & 33.1 & 25.7 & 42.6 \\
    $N(\rm H_{2}$) ($\times 10^{21}$ cm$^{-2}$) & 33.0(7.0) & 4.0(1.0) & 52.0(9.0) \\
    $n(\rm H_{2}$) ($\times 10^{5}$ cm$^{-3}$) &4.0(0.9) & 0.57(0.15) & 4.9(0.8) \\
    $\Delta v_{\rm NT}$ $^{c}$ (km\,s$^{-1}$) & 0.35(0.02) & 0.35(0.02)  & 0.35(0.02)  \\
    b ($\degree$) & 14.0(1.5) & 15.3(6.2) & 23.8(4.0) \\
    $\sigma_{\theta}$ ($\degree$) &10.1(1.1) &11.0(4.5) &17.6(3.0) \\
    B$_{\rm pos}$ ($\mu$G) &116(20) -- 257(42) &40(17) -- 88(38) &73(14) -- 162(32) \\
    $\mathcal{M}_{\rm A}$ & 0.4(0.04) -- 0.9(0.1) &0.4(0.2) -- 1.0(0.4)  &0.7(0.1) -- 1.6(0.3) \\
    $v_{\rm A}$ (km\,s$^{-1}$) & 0.3(0.04) -- 0.7(0.1) & 0.3(0.1) -- 0.7(0.3) & 0.2(0.03) -- 0.4(0.1) \\ 
    $\lambda$ & - & 0.3(0.2) -- 0.8(0.4) & 2.4(0.6) -- 5.4(1.4) \\
    $E_{\rm B}$ ($\times 10^{35}$ J) & 0.5(0.2) -- 2.6(0.9) & -- & 0.5(0.2) -- 2.2(0.9)\\ \hline
    \end{tabular}
    \label{tab:tab3}
    \end{center}
    \begin{center}
    a. Standard deviation of the Gaussian fit is in parentheses \\
    b. PA of ellipses is counter-clockwise from North \\
    c. $\Delta v_{\rm NT}$ values from \citet{Caselli_2002} \\
    \end{center}
	
\end{table*}
\end{center}

\section{Discussion}
\label{sec:disc}

\subsection{Contribution of the Magnetic Field}
\label{subsec:bfield_str}

The mass-to-flux ratio $\lambda$ is a parameter that can be used to quantify the importance of magnetic fields relative to gravity \citep{2004Ap&SS.292..225C}. It compares the critical value for the mass which can be supported by the magnetic flux $\Phi$, M$_{B_{\rm crit}}$ = $\Phi$/2$\pi\sqrt{G}$ \citep{1978PASJ...30..671N} to the observed mass and flux values. If the column density N and magnetic field strength B can be measured, the observed value for the ratio between mass and flux is $(M/\Phi)_{\rm obs}$=$mNA / BA$ and then the mass-to-flux ratio $\lambda$ is,

\begin{equation}
\lambda= \frac{(M/\Phi)_{\rm obs}}{(M/\Phi)_{\rm crit}} = 7.6\times10^{-21}\frac{N_{\rm H_{2}}(\rm cm^{-2})}{B_{\rm pos}(\mu \rm G)} \,
\label{eq:lam}
\end{equation}
\noindent
where $m$=2.8~$m_{\rm H}$, $N_{\rm H_2}$ is the molecular hydrogen column density and $B_{\rm pos}$ is the plane-of-sky magnetic field strength. When $\lambda < 1$, then the magnetic field is strong enough to support against gravity; this is referred as a \enquote{magnetically sub-critical} regime. Alternatively, if $\lambda > 1$, then the magnetic field is insufficient by itself to oppose gravity, and the cloud is instead \enquote{magnetically super-critical}. 

Using Equation\,\ref{eq:lam}, we calculate the mass-to-flux ratio to be 0.3--0.8 in the low-density periphery of the core, Region 1. We calculate a mass-to-flux ratio of 2.4--5.4 in the denser part of the core, Region 2. According to \citet{2004Ap&SS.292..225C} these ratios can be overestimated due to geometric biases and they suggest it can be overestimated by a factor up to 3, although it is a statistical correction and its application to individual measurements is unclear. If we consider this statistical correction, the mass-to-flux ratios are $\sim$0.1--0.3 and $\sim$0.8-1.8 in Regions 1 and 2 respectively. These are then the lower limits for the mass-to-flux ratio in each region.

We also get Alfv\'en Mach numbers of 0.4--1.0 and 0.7--1.6 for Regions 1 and 2 respectively, suggesting that both regions are roughly trans-critical and magnetic field and turbulence may play equal parts in support. As noted previously, the velocity information we are using is for the main starless core which is near Region 2, but further from Region 1 which is the low-density area on the periphery of the main dense core, and it does not resolve individual parts of the molecular cloud. We also find Alfv\'en velocities in the range of 0.2--0.7\,km\,s$^{-1}$ throughout the cloud.

For the lower-density Region 1, the region is entirely magnetically sub-critical indicating that the region may still be sufficiently supported against gravitational collapse by the magnetic field. It is also slightly sub-Alfv\'enic, meaning the magnetic field may play the dominant role. In the denser Region 2, we obtain a more definitively supercritical mass-to-flux ratio (it should be noted the large uncertainties with this value, as well as the results of the statistical correction), with a lower limit approaching trans- to sub-critical. This gradient of a sub-critical envelope (Region 1) transitioning into a trans- to super-critical core (Region 2) at sufficient densities is described in \citet{2004Ap&SS.292..225C}. It does suggest that the magnetic field is not sufficiently strong to support against gravitational collapse in the main starless core. However, it is worth mentioning that there are still many other processes in the molecular cloud such as turbulence and the influence of the CO outflow that could prevent gravitational collapse into a stellar object. While we don't yet see a protostar forming, as mentioned in Section~\ref{subsec:850dust}, we do see some possible fragmentation within the starless core where the densest part may be undergoing gravitational collapse.

\citet{2017ApJ...838...10M} suggested from modelling that L43 has formed all of the stars it will form in its lifetime and does not contain sufficient amounts of dense gas for further star formation. However we find higher column density values than they use for their modelling and do see potential fragmentation in the core. \citet{Chen_2009} finds that the main L43 starless core has observed DCO$^{+}$ and HCO$^{+}$ abundances that are higher and lower than modeled abundances respectively for an assumed amount of CO depletion. They suggest this indicates more CO depletion in the core and that the L43 starless core is spending a longer time at the higher density pre-protostellar core phase. If this is the case, additional supports such as turbulence may be needed to continue support against gravitational collapse in Region 2 since it seems to be moving beyond the stage where magnetic fields are critical. But findings of Region 2 in near equilibrium (within errors) also validates the long-lived age of the core that \citet{Chen_2009} finds. The local magnetic field appears to still be significant in the more diffuse Region 1, but this region is beyond the area considered by \citet{Chen_2009}. Additionally, we must consider that the CO outflow has potentially altered the structure of the magnetic field in the cloud, even within the starless core, potentially weakening or strengthening the field.

\subsection{Interaction of the Magnetic Field with Outflows}
\label{subsec:mag_co}

The spatial alignment (in the plane of sky) of the magnetic field in L43 with the outflow cavity walls can be seen in Figure\,\ref{fig:b_regions}, where the cyan outflow spatially overlaps with many of the magnetic field vectors. The magnetic field vectors that coincide with the CO outflow show a uniform distribution and strong peak at 146$\degree$. This coincides well with the outflow direction which we have taken to be $\sim$150$\pm$10$\degree$ due to it curving slightly.

\citet{1994ApJ...423..674W} suggested that RNO 91 sits in the foreground of the general L43 molecular cloud and so the possibility exists that there is in fact no physical association between the magnetic field and outflow. However, as was mentioned in Sec.~\ref{subsec:850dust} and as can be clearly seen in Figure~\ref{fig:herschel}, the outflow appears to have carved a cavity out of the molecular cloud, indicating it is to some degree embedded. This was also suggested by \citet{1988ApJ...330..385M}.

Alternatively, we may be tracing magnetic fields in the outflow cavity walls. In this case, as mentioned in Section\,\ref{subsec:cd_temp}, some of the Stokes \textit{I} emission we see, especially in the regions coincident with the outflow cavity walls, may be CO features \citep{2012MNRAS.426...23D}, especially the isolated emission to the south of the main cloud. We expect some contribution to the measured Stokes \textit{I} emission from CO but such contributions are typically less than \,20\% of the total emission observed \citep{2012MNRAS.426...23D,2015MNRAS.450.1094P,2016MNRAS.457.2139C} and in our case, we see contributions of $\sim$5--15\%. However, considering we do have some CO contribution, we cannot rule out the possibility that some fraction of the polarized emission in this region arises from CO polarization, polarized through the Goldreich-Kylafis effect \citep{1981ApJ...243L..75G,1982ApJ...253..606G}. This would add a further $\pm$\,90$\degree$ ambiguity on the magnetic field orientation.

On the other hand, we can assume the polarization and emission is not purely CO based as the emission features are also seen in all of the \textit{Herschel} bands and persist after CO subtraction in 850\,$\mu$m, indicating that there is a real dust feature present. A similar ``hollow shell'' morphology of dust emission in the presence of outflows is discussed in \citet{2006ApJ...645..357M}. Additionally, \citet{1998mnras...299..965} suggests that the mere presence of CO suggests some amount of dust shielding (from the UV field) in the outflow region. So we still consider it probable that we are tracing dust polarization in the outflow cavity walls.

Additionally, the relationship between magnetic fields with outflows has been observed on numerous occasions in other sources, on both JCMT and ALMA scales (see \citealt{2017ApJ...847...92H,Hull_2020,2021ApJ...918...85L,2022MNRAS.515.1026P}). \citet{2017ApJ...847...92H}, \citet{Hull_2020} and \citet{2022MNRAS.515.1026P} also see a similar alignment between the magnetic field and the cavity wall of the outflow to that which we see in L43/RNO 91. The benefit of our larger field of view here, when compared to ALMA, is that we can compare the magnetic field of the outflow to that in the surrounding regions and see that there is not just a preferential direction northwest to southeast, but rather that the magnetic field in the outflow region is actually different to that in the rest of the cloud. It should be noted that in regions observed by the JCMT, a preferred misalignment of 50$\degree$ between magnetic fields and outflows has previously been identified in a larger statistical sample \citep{2021ApJ...907...33Y}. The outflow observations in that study are largely on envelope- or small-scales (i.e. tens of arcseconds) rather than large-scale outflows like we see here. So while there is a statistically preferred misalignment between magnetic fields as observed by JCMT and outflows, we do see a clear indication of this occurring in L43 but rather see good alignment between outflow and magnetic fields. This is perhaps because we have such distinct large-scale cavity outflow walls which is what JCMT may be preferentially tracing. RNO 91 would also be interesting to follow up with ALMA polarization observations since there are smaller scale outflows in the envelope as well \citep{2005ApJ...624..841L,arce2006}. This could be more directly compared to observations by \citet{2017ApJ...847...92H} and \citet{Hull_2020} and to the statistical sample in \citet{2021ApJ...907...33Y}.

In RNO 91, the CO outflow was found to have a lower limit of its energy at $\approx 10^{29}$J \citep{1998mnras...299..965}. However this assumes a Class II source lifetime when considering how long the CO has been exposed to UV radiation, though a Class I lifetime would still be longer than the un-shielded CO lifetime of a few hundred years \citep{1998mnras...299..965}. So the CO outflow likely had a larger energy once and may have been able to influence the magnetic field orientation. Other studies have found the CO outflow energy to be $\sim5\times10^{35}$J \citep{myers1988} and $\sim1.4\times10^{35}$J \citep{arce2006}. These values are comparable to the magnetic energy values we see in L43, which is what we would expect. We can calculate the magnetic energy in Region 2 and the outflow region using

\begin{equation}
    E_{\rm B}(\rm J)=10^{-20}\frac{B^2(\rm \mu G ^2) V(\rm m^{3})}{2\mu_o(\rm N\,A^{-2})} \,
    \label{eq:eb}
\end{equation}
\noindent
where $B$ is the magnetic field strength in $\mu$G (as measured in the plane-of-sky), $V$ is the volume of the region in m$^{3}$ and $\mu_{o}$ is the permeability of free space, giving us the magnetic energy in Joules. Since we consider the outflow to have affected the dense gas and dust and dragged the magnetic field, we would expect the outflow energy to be at least equal to the magnetic energy. Assuming an ellipsoid shape when calculating the volume of both regions (see Table~\ref{tab:tab3} for ellipse parameters), we find magnetic energies of $\approx$0.5--2.5$\times$10$^{35}$J. This suggests that we can help further place a lower limit on the outflow energy of $\approx$0.5--2.5$\times$10$^{35}$J, which is comparable to the values stated above, though we do remain cautious of the magnetic field strength derived in the outflow region as mentioned before.

\subsection{Evolution of this isolated filament}
\label{subsec:filevo}

One point of interest in this region is that there is a very clear evolutionary gradient from southwest to northeast. Initially there is the evolved T-Tauri star RNO 90 which is the oldest source and currently has no known large-scale outflows. It has also formed a protostellar disk \citep{pont2010}. RNO 91, which is further along the filament, is a protostellar source which drives the now familiar CO outflow. Then finally the starless core sits $\approx$10,000\,AU further along the filament. The orientation of this filament is such that it extends roughly radially away from Sco OB2, with RNO 90 the closest to Sco OB2. The filament sits $\approx$42\,pc away from Sco OB2 (in the plane-of-sky) assuming a general distance of 125\,pc. While this may be merely a coincidence, it is interesting that the evolutionary track in such an isolated filament starts in the part of the filament pointing directly towards Sco OB2 (in the plane-of-sky). The Ophiuchus region and star formation within has previously been suggested to be shaped and driven by Sco OB2 \citep{loren1989}.

Additionally, we can picture two evolutionary scenarios for the starless core, scenarios that with present observations we cannot distinguish between and that may very well be happening at the same time. The starless core L43 has formed with its long axis parallel to the outflow cavity wall. This could suggest that material has been funnelled down the filament which is also parallel with the large-scale Planck field and is building up on the outflow cavity walls. Build-up has not occurred so readily along the western wall of the outflow cavity because there is less material available for accretion to the west since RNO 90 has already been formed. However, it could also be the case that the dense core already existed and fragmented to form both the starless core and RNO 91 and the starless core has no been compressed along the filament orientation by the outflow. It is difficult to differentiate between these two scenarios and the possibility of course remains that they could both be true, with an initial fragmentation that has become denser over time. \citet{2020ApJS..249...33K} does suggest that the starless core is a `late' or chemically-evolved, starless core as determined by a high $N$(DNC)/$N$(HN$^{13}$C) ratio and line detection in N$_{2}$D$^{+}$. So the core may have formed at a similar time to RNO 91 but has since had its evolution slightly delayed due to injected turbulence by RNO 91 as well as less readily available material to form a star with.

\section{Summary}
\label{sec:summary}

We presented polarization measurements of the infrared dark molecular cloud L43 at 850\,$\mu$m made using JCMT/POL-2 as part of the JCMT BISTRO Survey. We found H$_{2}$ column densities on the order of 10$^{22}$-10$^{23}$\,cm$^{-2}$, which are typical values in dense starless cores. We measured a power law index of $\sim$ -0.85 when plotting polarization percentage as a function of total 850\,$\mu$m intensity, indicating a possible decrease, but not complete loss, in grain alignment efficiency, deep within the molecular cloud. By rotating the polarization vectors by 90$^{\circ}$, we inferred the magnetic field orientation in L43 and saw a complicated and multiple-component magnetic field. 

We divided the magnetic field into three regions, with one region slightly offset from the dense submillimetre-bright core (Region 2), another region in the more diffuse region to the east (Region 1) and then vectors which spatially coincides in the plane of the sky with the CO outflow driven by RNO 91. We saw alignment between the magnetic field and the outflow cavity walls which is distinctly different from the magnetic field in the rest of the cloud. We calculated the magnetic field strengths of $\sim$40$\pm$20--90$\pm$40\,$\mu$G in Region 1 and $\sim$70$\pm$15--160$\pm$30\,$\mu$G in Region 2. We did calculate a magnetic field strength in the outflow region of $\sim$120$\pm$20--260$\pm$40\,$\mu$G but advise caution with interpreting this value. Region 1 appeared to be magnetically sub- or trans-critical and but sub-Alfv\'enic. This suggested that the magnetic field is still important in comparison to gravity and turbulent motions. Region 2 is both magnetically super-critical and sub- to trans-Alfv\'enic so the magnetic field may not be playing a significant role. This is compounded by potential fragmentation in the main core, suggesting it could be heading towards forming a protostar. We also proposed an evolutionary gradient across the isolated filament starting with the most evolved source RNO 90 which is closest to the Sco OB2 association and moving away from Sco OB2 towards RNO 91 and then eventually the starless core.


%

\bigbreak
\bigbreak

We thank the referee for their time and constructive feedback to help improve this article. J.K. acknowledges funding from the Moses Holden Studentship for his PhD. D.W.-T. acknowledge Science and Technology Facilities Council (STFC) support under grant number ST\textbackslash R000786\textbackslash1. K.P. is a Royal Society University Research Fellow, supported by grant number URF\textbackslash R1\textbackslash211322. C.W.L. acknowledges the support by the BasicScience Research Program through the National Research Foundation of Korea (NRF) funded by the Ministry of Education, Science and Technology (NRF-2019R1A2C1010851), and the support by the Korea Astronomy and Space Science Institute grant funded by the Korea government (MSIT) (Project No. 2022-1-840-05). W.K. was supported by the National Research Foundation of Korea (NRF) grant funded by the Korea government (MSIT) (NRF-2021R1F1A1061794). K.Q. is supported by National Key R\&D Program of China No. 2022YFA1603100, No. 2017YFA0402604, the National Natural Science Foundation of China (NSFC) grant U1731237, and the science research grant from the China Manned Space Project with No. CMS-CSST-2021-B06. C.L.H.H. acknowledges the support of the NAOJ Fellowship and JSPS KAKENHI grants KG18K13586, KG20K14527, and KG22H01271. M.R. is supported by the international Gemini Observatory, a program of NSF’s NOIRLab, which is managed by the Association of Universities for Research in Astronomy (AURA) under a cooperative agreement with the National Science Foundation, on behalf of the Gemini partnership of Argentina, Brazil, Canada, Chile, the Republic of Korea, and the United States of America. L.F. acknowledges support from the Ministry of Science and Technology of Taiwan, under grants number 111-2811-M-005 -007 and 109-2112-M-005 -003 -MY3. M.T. is supported by JSPS KAKENHI grant Nos.18H05442,15H02063,and 22000005. J.K.is supported by JSPS KAKENHI grant No.19K14775. D.J. is supported by NRC Canada and by an NSERC Discovery Grant. E.J.C. is supported by Basic Science Research Program through the National Research Foundation of Korea(NRF) funded by the Ministry of Education(grant number) (NRF-2022R1I1A1A01053862). F.P. acknowledges support from the Spanish State Research Agency (AEI) under grant number PID2019-105552RB-C43 and from the Agencia Canaria de Investigación, Innovación y Sociedad de la Información (ACIISI) under the European FEDER (FONDO EUROPEO DE DESARROLLO REGIONAL) de Canarias 2014-2020 grant No. PROID2021010078. C.E. acknowledges the financial support from grant RJF/2020/000071 as a part of Ramanujan Fellowship awarded by Science and Engineering Research Board (SERB), Department of Science and Technology (DST), Govt. of India. The James Clerk Maxwell Telescope is operated by the East Asian Observatory on behalf of The National Astronomical Observatory of Japan; Academia Sinica Institute of Astronomy and Astrophysics; the Korea Astronomy and Space Science Institute; the National Astronomical Research Institute of Thailand; Center for Astronomical Mega-Science (as well as the National Key R\&D Program of China with No. 2017YFA0402700). Additional funding support is provided by the Science and Technology Facilities Council of the United Kingdom and participating universities and organizations in the United Kingdom, Canada and Ireland. The authors wish to recognize and acknowledge the very significant cultural role and reverence that the summit of Maunakea has always had within the indigenous Hawaiian community.  We are most fortunate to have the opportunity to conduct observations from this mountain. The data taken in this paper were observed under the project code M20AL018. The Starlink software \citep{2014ASPC..485..391C} is currently supported by the East Asian Observatory.

\vspace{5mm}
\facilities{JCMT (SCUBA-2, POL-2), Herschel (SPIRE, PACS)}

\software{Starlink \citep{2014ASPC..485..391C}, Astropy \citep{2013A&A...558A..33A,2018AJ....156..123A}}

\bibliography{citations}{}

\begin{thebibliography}{}
\expandafter\ifx\csname natexlab\endcsname\relax\def\natexlab#1{#1}\fi
\providecommand{\url}[1]{\href{#1}{#1}}
\providecommand{\dodoi}[1]{doi:~\href{http://doi.org/#1}{\nolinkurl{#1}}}
\providecommand{\doeprint}[1]{\href{http://ascl.net/#1}{\nolinkurl{http://ascl.net/#1}}}
\providecommand{\doarXiv}[1]{\href{https://arxiv.org/abs/#1}{\nolinkurl{https://arxiv.org/abs/#1}}}

\bibitem[{{Andersson} {et~al.}(2015){Andersson}, {Lazarian}, \&
  {Vaillancourt}}]{ALV2015}
{Andersson}, B.~G., {Lazarian}, A., \& {Vaillancourt}, J.~E. 2015, \araa, 53,
  501, \dodoi{10.1146/annurev-astro-082214-122414}

\bibitem[{{Andre} \& {Montmerle}(1994)}]{1994ApJ...420..837A}
{Andre}, P., \& {Montmerle}, T. 1994, \apj, 420, 837, \dodoi{10.1086/173608}

\bibitem[{{Arce} \& {Sargent}(2006)}]{arce2006}
{Arce}, H.~G., \& {Sargent}, A.~I. 2006, \apj, 646, 1070,
  \dodoi{10.1086/505104}

\bibitem[{{Arzoumanian} {et~al.}(2021){Arzoumanian}, {Furuya}, {Hasegawa},
  {Tahani}, {Sadavoy}, {Hull}, {Johnstone}, {Koch}, {Inutsuka}, {Doi}, {Hoang},
  {Onaka}, {Iwasaki}, {Shimajiri}, {Inoue}, {Peretto}, {Andr{\'e}}, {Bastien},
  {Berry}, {Chen}, {Di Francesco}, {Eswaraiah}, {Fanciullo}, {Fissel}, {Hwang},
  {Kang}, {Kim}, {Kim}, {Kirchschlager}, {Kwon}, {Lee}, {Liu}, {Lyo}, {Pattle},
  {Soam}, {Tang}, {Whitworth}, {Ching}, {Coud{\'e}}, {Wang}, {Ward-Thompson},
  {Lai}, {Qiu}, {Bourke}, {Byun}, {Chen}, {Chen}, {Chen}, {Cho}, {Choi},
  {Choi}, {Chrysostomou}, {Chung}, {Dai}, {Diep}, {Duan}, {Duan}, {Eden},
  {Fiege}, {Franzmann}, {Friberg}, {Fuller}, {Gledhill}, {Graves}, {Greaves},
  {Griffin}, {Gu}, {Han}, {Hatchell}, {Hayashi}, {Houde}, {Jeong}, {Kang},
  {Kang}, {Kataoka}, {Kawabata}, {Kemper}, {Kim}, {Kim}, {Kim}, {Kim}, {Kirk},
  {Kobayashi}, {K{\"o}nyves}, {Kusune}, {Kwon}, {Lacaille}, {Law}, {Lee},
  {Lee}, {Lee}, {Lee}, {Lee}, {Li}, {Li}, {Li}, {Liu}, {Liu}, {Liu}, {Lu},
  {Mairs}, {Matsumura}, {Matthews}, {Moriarty-Schieven}, {Nagata}, {Nakamura},
  {Nakanishi}, {Ngoc}, {Ohashi}, {Park}, {Parsons}, {Pyo}, {Qian}, {Rao},
  {Rawlings}, {Rawlings}, {Retter}, {Richer}, {Rigby}, {Saito}, {Savini},
  {Scaife}, {Seta}, {Shinnaga}, {Tamura}, {Tang}, {Tomisaka}, {Tram},
  {Tsukamoto}, {Viti}, {Wang}, {Xie}, {Yen}, {Yoo}, {Yuan}, {Yun}, {Zenko},
  {Zhang}, {Zhang}, {Zhang}, {Zhou}, {Zhu}, {de Looze}, {Dowell}, {Eyres},
  {Falle}, {Friesen}, {Robitaille}, \& {van Loo}}]{2021A&A...647A..78A}
{Arzoumanian}, D., {Furuya}, R.~S., {Hasegawa}, T., {et~al.} 2021, \aap, 647,
  A78, \dodoi{10.1051/0004-6361/202038624}

\bibitem[{{Astropy Collaboration} {et~al.}(2013){Astropy Collaboration},
  {Robitaille}, {Tollerud}, {Greenfield}, {Droettboom}, {Bray}, {Aldcroft},
  {Davis}, {Ginsburg}, {Price-Whelan}, {Kerzendorf}, {Conley}, {Crighton},
  {Barbary}, {Muna}, {Ferguson}, {Grollier}, {Parikh}, {Nair}, {Unther},
  {Deil}, {Woillez}, {Conseil}, {Kramer}, {Turner}, {Singer}, {Fox}, {Weaver},
  {Zabalza}, {Edwards}, {Azalee Bostroem}, {Burke}, {Casey}, {Crawford},
  {Dencheva}, {Ely}, {Jenness}, {Labrie}, {Lim}, {Pierfederici}, {Pontzen},
  {Ptak}, {Refsdal}, {Servillat}, \& {Streicher}}]{2013A&A...558A..33A}
{Astropy Collaboration}, {Robitaille}, T.~P., {Tollerud}, E.~J., {et~al.} 2013,
  \aap, 558, A33, \dodoi{10.1051/0004-6361/201322068}

\bibitem[{{Astropy Collaboration} {et~al.}(2018){Astropy Collaboration},
  {Price-Whelan}, {Sip{\H{o}}cz}, {G{\"u}nther}, {Lim}, {Crawford}, {Conseil},
  {Shupe}, {Craig}, {Dencheva}, {Ginsburg}, {VanderPlas}, {Bradley},
  {P{\'e}rez-Su{\'a}rez}, {de Val-Borro}, {Aldcroft}, {Cruz}, {Robitaille},
  {Tollerud}, {Ardelean}, {Babej}, {Bach}, {Bachetti}, {Bakanov}, {Bamford},
  {Barentsen}, {Barmby}, {Baumbach}, {Berry}, {Biscani}, {Boquien}, {Bostroem},
  {Bouma}, {Brammer}, {Bray}, {Breytenbach}, {Buddelmeijer}, {Burke},
  {Calderone}, {Cano Rodr{\'\i}guez}, {Cara}, {Cardoso}, {Cheedella}, {Copin},
  {Corrales}, {Crichton}, {D'Avella}, {Deil}, {Depagne}, {Dietrich}, {Donath},
  {Droettboom}, {Earl}, {Erben}, {Fabbro}, {Ferreira}, {Finethy}, {Fox},
  {Garrison}, {Gibbons}, {Goldstein}, {Gommers}, {Greco}, {Greenfield},
  {Groener}, {Grollier}, {Hagen}, {Hirst}, {Homeier}, {Horton}, {Hosseinzadeh},
  {Hu}, {Hunkeler}, {Ivezi{\'c}}, {Jain}, {Jenness}, {Kanarek}, {Kendrew},
  {Kern}, {Kerzendorf}, {Khvalko}, {King}, {Kirkby}, {Kulkarni}, {Kumar},
  {Lee}, {Lenz}, {Littlefair}, {Ma}, {Macleod}, {Mastropietro}, {McCully},
  {Montagnac}, {Morris}, {Mueller}, {Mumford}, {Muna}, {Murphy}, {Nelson},
  {Nguyen}, {Ninan}, {N{\"o}the}, {Ogaz}, {Oh}, {Parejko}, {Parley}, {Pascual},
  {Patil}, {Patil}, {Plunkett}, {Prochaska}, {Rastogi}, {Reddy Janga},
  {Sabater}, {Sakurikar}, {Seifert}, {Sherbert}, {Sherwood-Taylor}, {Shih},
  {Sick}, {Silbiger}, {Singanamalla}, {Singer}, {Sladen}, {Sooley},
  {Sornarajah}, {Streicher}, {Teuben}, {Thomas}, {Tremblay}, {Turner},
  {Terr{\'o}n}, {van Kerkwijk}, {de la Vega}, {Watkins}, {Weaver}, {Whitmore},
  {Woillez}, {Zabalza}, \& {Astropy Contributors}}]{2018AJ....156..123A}
{Astropy Collaboration}, {Price-Whelan}, A.~M., {Sip{\H{o}}cz}, B.~M., {et~al.}
  2018, \aj, 156, 123, \dodoi{10.3847/1538-3881/aabc4f}

\bibitem[{{Bailer-Jones} {et~al.}(2018){Bailer-Jones}, {Rybizki}, {Fouesneau},
  {Mantelet}, \& {Andrae}}]{bailerjones18}
{Bailer-Jones}, C.~A.~L., {Rybizki}, J., {Fouesneau}, M., {Mantelet}, G., \&
  {Andrae}, R. 2018, \aj, 156, 58, \dodoi{10.3847/1538-3881/aacb21}

\bibitem[{{Beckwith} {et~al.}(1990){Beckwith}, {Sargent}, {Chini}, \&
  {Guesten}}]{1990AJ.....99..924B}
{Beckwith}, S. V.~W., {Sargent}, A.~I., {Chini}, R.~S., \& {Guesten}, R. 1990,
  \aj, 99, 924, \dodoi{10.1086/115385}

\bibitem[{Bence {et~al.}(1998)Bence, Padman, Isaak, Wiedner, \&
  Wright}]{1998mnras...299..965}
Bence, S., Padman, R., Isaak, K., Wiedner, M., \& Wright, G. 1998, \mnras, 299,
  965, \dodoi{10.1046/j.1365-8711.1998.01789.x}

\bibitem[{{Bonnor}(1956)}]{1956MNRAS.116..351B}
{Bonnor}, W.~B. 1956, \mnras, 116, 351, \dodoi{10.1093/mnras/116.3.351}

\bibitem[{{Caselli} {et~al.}(2002){Caselli}, {Benson}, {Myers}, \&
  {Tafalla}}]{Caselli_2002}
{Caselli}, P., {Benson}, P.~J., {Myers}, P.~C., \& {Tafalla}, M. 2002, \apj,
  572, 238, \dodoi{10.1086/340195}

\bibitem[{{Chandrasekhar} \& {Fermi}(1953)}]{1953ApJ...118..116C}
{Chandrasekhar}, S., \& {Fermi}, E. 1953, \apj, 118, 116,
  \dodoi{10.1086/145732}

\bibitem[{{Chapin} {et~al.}(2013){Chapin}, {Berry}, {Gibb}, {Jenness}, {Scott},
  {Tilanus}, {Economou}, \& {Holland}}]{2013MNRAS.430.2545C}
{Chapin}, E.~L., {Berry}, D.~S., {Gibb}, A.~G., {et~al.} 2013, \mnras, 430,
  2545, \dodoi{10.1093/mnras/stt052}

\bibitem[{Chen {et~al.}(2009)Chen, Evans, Lee, \& Bourke}]{Chen_2009}
Chen, J.-H., Evans, N.~J., Lee, J.-E., \& Bourke, T.~L. 2009, \apj, 705, 1160,
  \dodoi{10.1088/0004-637x/705/2/1160}

\bibitem[{{Cohen}(1980)}]{1980AJ.....85...29C}
{Cohen}, M. 1980, \aj, 85, 29, \dodoi{10.1086/112630}

\bibitem[{{Coud{\'e}} {et~al.}(2016){Coud{\'e}}, {Bastien}, {Kirk},
  {Johnstone}, {Drabek-Maunder}, {Graves}, {Hatchell}, {Chapin}, {Gibb},
  {Matthews}, \& {JCMT Gould Belt Survey Team}}]{2016MNRAS.457.2139C}
{Coud{\'e}}, S., {Bastien}, P., {Kirk}, H., {et~al.} 2016, \mnras, 457, 2139,
  \dodoi{10.1093/mnras/stv3009}

\bibitem[{{Crutcher}(2004)}]{2004Ap&SS.292..225C}
{Crutcher}, R.~M. 2004, \apss, 292, 225,
  \dodoi{10.1023/B:ASTR.0000045021.42255.95}

\bibitem[{{Crutcher} {et~al.}(2004){Crutcher}, {Nutter}, {Ward-Thompson}, \&
  {Kirk}}]{2004ApJ...600..279C}
{Crutcher}, R.~M., {Nutter}, D.~J., {Ward-Thompson}, D., \& {Kirk}, J.~M. 2004,
  \apj, 600, 279, \dodoi{10.1086/379705}

\bibitem[{{Currie} \& {Berry}(2014)}]{2014ascl.soft03022C}
{Currie}, M.~J., \& {Berry}, D.~S. 2014, {KAPPA: Kernel Applications Package}.
\newblock \doeprint{1403.022}

\bibitem[{{Currie} {et~al.}(2014){Currie}, {Berry}, {Jenness}, {Gibb}, {Bell},
  \& {Draper}}]{2014ASPC..485..391C}
{Currie}, M.~J., {Berry}, D.~S., {Jenness}, T., {et~al.} 2014, Astronomical
  Society of the Pacific Conference Series, Vol. 485, {Starlink Software in
  2013}, ed. N.~{Manset} \& P.~{Forshay} (Astronomical Society of the Pacific),
  391

\bibitem[{{Davis}(1951)}]{1951PhRv...81..890D}
{Davis}, L. 1951, \physrev, 81, 890, \dodoi{10.1103/PhysRev.81.890.2}

\bibitem[{{de Geus} {et~al.}(1990){de Geus}, {Bronfman}, \&
  {Thaddeus}}]{1990A&A...231..137D}
{de Geus}, E.~J., {Bronfman}, L., \& {Thaddeus}, P. 1990, \aap, 231, 137

\bibitem[{{Dempsey} {et~al.}(2013){Dempsey}, {Friberg}, {Jenness}, {Tilanus},
  {Thomas}, {Holland}, {Bintley}, {Berry}, {Chapin}, {Chrysostomou}, {Davis},
  {Gibb}, {Parsons}, \& {Robson}}]{2013MNRAS.430.2534D}
{Dempsey}, J.~T., {Friberg}, P., {Jenness}, T., {et~al.} 2013, \mnras, 430,
  2534, \dodoi{10.1093/mnras/stt090}

\bibitem[{{Drabek} {et~al.}(2012){Drabek}, {Hatchell}, {Friberg}, {Richer},
  {Graves}, {Buckle}, {Nutter}, {Johnstone}, \& {Di
  Francesco}}]{2012MNRAS.426...23D}
{Drabek}, E., {Hatchell}, J., {Friberg}, P., {et~al.} 2012, \mnras, 426, 23,
  \dodoi{10.1111/j.1365-2966.2012.21140.x}

\bibitem[{{Ebert}(1955)}]{1955ZA.....37..217E}
{Ebert}, R. 1955, \zap, 37, 217

\bibitem[{Federrath(2015)}]{fed2015}
Federrath, C. 2015, \mnras, 450, 4035, \dodoi{10.1093/mnras/stv941}

\bibitem[{{Feh{\'e}r} {et~al.}(2022){Feh{\'e}r}, {T{\'o}th}, {Kraus},
  {B{\H{o}}gner}, {Kim}, {Liu}, {Tatematsu}, {T{\'o}th}, {Eden}, {Hirano},
  {Juvela}, {Kim}, {Li}, {Liu}, {Wu}, {Wu}, \& {Top-Scope
  Collaboration}}]{feher2022}
{Feh{\'e}r}, O., {T{\'o}th}, L.~V., {Kraus}, A., {et~al.} 2022, \apjs, 258, 17,
  \dodoi{10.3847/1538-4365/ac3337}

\bibitem[{{Friberg} {et~al.}(2016){Friberg}, {Bastien}, {Berry}, {Savini},
  {Graves}, \& {Pattle}}]{2016SPIE.9914E..03F}
{Friberg}, P., {Bastien}, P., {Berry}, D., {et~al.} 2016, Society of
  Photo-Optical Instrumentation Engineers (SPIE) Conference Series, Vol. 9914,
  {POL-2: a polarimeter for the James-Clerk-Maxwell telescope} (Society of
  Photo-Optical Instrumentation Engineers), 991403, \dodoi{10.1117/12.2231943}

\bibitem[{{Gaia Collaboration} {et~al.}(2021){Gaia Collaboration}, {Brown},
  {Vallenari}, {Prusti}, {de Bruijne}, {Babusiaux}, {Biermann}, {Creevey},
  {Evans}, {Eyer}, {Hutton}, {Jansen}, {Jordi}, {Klioner}, {Lammers},
  {Lindegren}, {Luri}, {Mignard}, {Panem}, {Pourbaix}, {Randich}, {Sartoretti},
  {Soubiran}, {Walton}, {Arenou}, {Bailer-Jones}, {Bastian}, {Cropper},
  {Drimmel}, {Katz}, {Lattanzi}, {van Leeuwen}, {Bakker}, {Cacciari},
  {Casta{\~n}eda}, {De Angeli}, {Ducourant}, {Fabricius}, {Fouesneau},
  {Fr{\'e}mat}, {Guerra}, {Guerrier}, {Guiraud}, {Jean-Antoine Piccolo},
  {Masana}, {Messineo}, {Mowlavi}, {Nicolas}, {Nienartowicz}, {Pailler},
  {Panuzzo}, {Riclet}, {Roux}, {Seabroke}, {Sordo}, {Tanga}, {Th{\'e}venin},
  {Gracia-Abril}, {Portell}, {Teyssier}, {Altmann}, {Andrae}, {Bellas-Velidis},
  {Benson}, {Berthier}, {Blomme}, {Brugaletta}, {Burgess}, {Busso}, {Carry},
  {Cellino}, {Cheek}, {Clementini}, {Damerdji}, {Davidson}, {Delchambre},
  {Dell'Oro}, {Fern{\'a}ndez-Hern{\'a}ndez}, {Galluccio}, {Garc{\'\i}a-Lario},
  {Garcia-Reinaldos}, {Gonz{\'a}lez-N{\'u}{\~n}ez}, {Gosset}, {Haigron},
  {Halbwachs}, {Hambly}, {Harrison}, {Hatzidimitriou}, {Heiter},
  {Hern{\'a}ndez}, {Hestroffer}, {Hodgkin}, {Holl}, {Jan{\ss}en}, {Jevardat de
  Fombelle}, {Jordan}, {Krone-Martins}, {Lanzafame}, {L{\"o}ffler}, {Lorca},
  {Manteiga}, {Marchal}, {Marrese}, {Moitinho}, {Mora}, {Muinonen}, {Osborne},
  {Pancino}, {Pauwels}, {Petit}, {Recio-Blanco}, {Richards}, {Riello},
  {Rimoldini}, {Robin}, {Roegiers}, {Rybizki}, {Sarro}, {Siopis}, {Smith},
  {Sozzetti}, {Ulla}, {Utrilla}, {van Leeuwen}, {van Reeven}, {Abbas}, {Abreu
  Aramburu}, {Accart}, {Aerts}, {Aguado}, {Ajaj}, {Altavilla}, {{\'A}lvarez},
  {{\'A}lvarez Cid-Fuentes}, {Alves}, {Anderson}, {Anglada Varela}, {Antoja},
  {Audard}, {Baines}, {Baker}, {Balaguer-N{\'u}{\~n}ez}, {Balbinot}, {Balog},
  {Barache}, {Barbato}, {Barros}, {Barstow}, {Bartolom{\'e}}, {Bassilana},
  {Bauchet}, {Baudesson-Stella}, {Becciani}, {Bellazzini}, {Bernet}, {Bertone},
  {Bianchi}, {Blanco-Cuaresma}, {Boch}, {Bombrun}, {Bossini}, {Bouquillon},
  {Bragaglia}, {Bramante}, {Breedt}, {Bressan}, {Brouillet}, {Bucciarelli},
  {Burlacu}, {Busonero}, {Butkevich}, {Buzzi}, {Caffau}, {Cancelliere},
  {C{\'a}novas}, {Cantat-Gaudin}, {Carballo}, {Carlucci}, {Carnerero},
  {Carrasco}, {Casamiquela}, {Castellani}, {Castro-Ginard}, {Castro Sampol},
  {Chaoul}, {Charlot}, {Chemin}, {Chiavassa}, {Cioni}, {Comoretto}, {Cooper},
  {Cornez}, {Cowell}, {Crifo}, {Crosta}, {Crowley}, {Dafonte}, {Dapergolas},
  {David}, {David}, {de Laverny}, {De Luise}, {De March}, {De Ridder}, {de
  Souza}, {de Teodoro}, {de Torres}, {del Peloso}, {del Pozo}, {Delbo},
  {Delgado}, {Delgado}, {Delisle}, {Di Matteo}, {Diakite}, {Diener},
  {Distefano}, {Dolding}, {Eappachen}, {Edvardsson}, {Enke}, {Esquej}, {Fabre},
  {Fabrizio}, {Faigler}, {Fedorets}, {Fernique}, {Fienga}, {Figueras},
  {Fouron}, {Fragkoudi}, {Fraile}, {Franke}, {Gai}, {Garabato},
  {Garcia-Gutierrez}, {Garc{\'\i}a-Torres}, {Garofalo}, {Gavras}, {Gerlach},
  {Geyer}, {Giacobbe}, {Gilmore}, {Girona}, {Giuffrida}, {Gomel}, {Gomez},
  {Gonzalez-Santamaria}, {Gonz{\'a}lez-Vidal}, {Granvik},
  {Guti{\'e}rrez-S{\'a}nchez}, {Guy}, {Hauser}, {Haywood}, {Helmi}, {Hidalgo},
  {Hilger}, {H{\l}adczuk}, {Hobbs}, {Holland}, {Huckle}, {Jasniewicz},
  {Jonker}, {Juaristi Campillo}, {Julbe}, {Karbevska}, {Kervella}, {Khanna},
  {Kochoska}, {Kontizas}, {Kordopatis}, {Korn}, {Kostrzewa-Rutkowska},
  {Kruszy{\'n}ska}, {Lambert}, {Lanza}, {Lasne}, {Le Campion}, {Le Fustec},
  {Lebreton}, {Lebzelter}, {Leccia}, {Leclerc}, {Lecoeur-Taibi}, {Liao},
  {Licata}, {Lindstr{\o}m}, {Lister}, {Livanou}, {Lobel}, {Madrero Pardo},
  {Managau}, {Mann}, {Marchant}, {Marconi}, {Marcos Santos}, {Marinoni},
  {Marocco}, {Marshall}, {Martin Polo}, {Mart{\'\i}n-Fleitas}, {Masip},
  {Massari}, {Mastrobuono-Battisti}, {Mazeh}, {McMillan}, {Messina},
  {Michalik}, {Millar}, {Mints}, {Molina}, {Molinaro}, {Moln{\'a}r},
  {Montegriffo}, {Mor}, {Morbidelli}, {Morel}, {Morris}, {Mulone}, {Munoz},
  {Muraveva}, {Murphy}, {Musella}, {Noval}, {Ord{\'e}novic}, {Orr{\`u}},
  {Osinde}, {Pagani}, {Pagano}, {Palaversa}, {Palicio}, {Panahi}, {Pawlak},
  {Pe{\~n}alosa Esteller}, {Penttil{\"a}}, {Piersimoni}, {Pineau}, {Plachy},
  {Plum}, {Poggio}, {Poretti}, {Poujoulet}, {Pr{\v{s}}a}, {Pulone}, {Racero},
  {Ragaini}, {Rainer}, {Raiteri}, {Rambaux}, {Ramos}, {Ramos-Lerate}, {Re
  Fiorentin}, {Regibo}, {Reyl{\'e}}, {Ripepi}, {Riva}, {Rixon}, {Robichon},
  {Robin}, {Roelens}, {Rohrbasser}, {Romero-G{\'o}mez}, {Rowell}, {Royer},
  {Rybicki}, {Sadowski}, {Sagrist{\`a} Sell{\'e}s}, {Sahlmann}, {Salgado},
  {Salguero}, {Samaras}, {Sanchez Gimenez}, {Sanna}, {Santove{\~n}a},
  {Sarasso}, {Schultheis}, {Sciacca}, {Segol}, {Segovia}, {S{\'e}gransan},
  {Semeux}, {Shahaf}, {Siddiqui}, {Siebert}, {Siltala}, {Slezak}, {Smart},
  {Solano}, {Solitro}, {Souami}, {Souchay}, {Spagna}, {Spoto}, {Steele},
  {Steidelm{\"u}ller}, {Stephenson}, {S{\"u}veges}, {Szabados}, {Szegedi-Elek},
  {Taris}, {Tauran}, {Taylor}, {Teixeira}, {Thuillot}, {Tonello}, {Torra},
  {Torra}, {Turon}, {Unger}, {Vaillant}, {van Dillen}, {Vanel}, {Vecchiato},
  {Viala}, {Vicente}, {Voutsinas}, {Weiler}, {Wevers}, {Wyrzykowski}, {Yoldas},
  {Yvard}, {Zhao}, {Zorec}, {Zucker}, {Zurbach}, \& {Zwitter}}]{gaia21}
{Gaia Collaboration}, {Brown}, A.~G.~A., {Vallenari}, A., {et~al.} 2021, \aap,
  649, A1, \dodoi{10.1051/0004-6361/202039657}

\bibitem[{{Garufi} {et~al.}(2022){Garufi}, {Dominik}, {Ginski}, {Benisty}, {van
  Holstein}, {Henning}, {Pawellek}, {Pinte}, {Avenhaus}, {Facchini},
  {Galicher}, {Gratton}, {M{\'e}nard}, {Muro-Arena}, {Milli}, {Stolker},
  {Vigan}, {Villenave}, {Moulin}, {Origne}, {Rigal}, {Sauvage}, \&
  {Weber}}]{garufi22}
{Garufi}, A., {Dominik}, C., {Ginski}, C., {et~al.} 2022, \aap, 658, A137,
  \dodoi{10.1051/0004-6361/202141692}

\bibitem[{{Goldreich} \& {Kylafis}(1981)}]{1981ApJ...243L..75G}
{Goldreich}, P., \& {Kylafis}, N.~D. 1981, \apjl, 243, L75,
  \dodoi{10.1086/183446}

\bibitem[{{Goldreich} \& {Kylafis}(1982)}]{1982ApJ...253..606G}
---. 1982, \apj, 253, 606, \dodoi{10.1086/159663}

\bibitem[{{Heiles} \& {Troland}(2005)}]{heilestrol05}
{Heiles}, C., \& {Troland}, T.~H. 2005, \apj, 624, 773, \dodoi{10.1086/428896}

\bibitem[{{Hennebelle} \& {Inutsuka}(2019)}]{2019FrASS...6....5H}
{Hennebelle}, P., \& {Inutsuka}, S.-i. 2019, \frass, 6, 5,
  \dodoi{10.3389/fspas.2019.00005}

\bibitem[{{Herbst} \& {Warner}(1981)}]{1981AJ.....86..885H}
{Herbst}, W., \& {Warner}, J.~W. 1981, \aj, 86, 885, \dodoi{10.1086/112962}

\bibitem[{{Hildebrand}(1983)}]{1983QJRAS..24..267H}
{Hildebrand}, R.~H. 1983, \qjras, 24, 267

\bibitem[{{Hildebrand} {et~al.}(2009){Hildebrand}, {Kirby}, {Dotson}, {Houde},
  \& {Vaillancourt}}]{hild09}
{Hildebrand}, R.~H., {Kirby}, L., {Dotson}, J.~L., {Houde}, M., \&
  {Vaillancourt}, J.~E. 2009, \apj, 696, 567,
  \dodoi{10.1088/0004-637X/696/1/567}

\bibitem[{{Hodapp}(1994)}]{1994ApJS...94..615H}
{Hodapp}, K.-W. 1994, \apjs, 94, 615, \dodoi{10.1086/192084}

\bibitem[{{Holland} {et~al.}(2013){Holland}, {Bintley}, {Chapin},
  {Chrysostomou}, {Davis}, {Dempsey}, {Duncan}, {Fich}, {Friberg}, {Halpern},
  {Irwin}, {Jenness}, {Kelly}, {MacIntosh}, {Robson}, {Scott}, {Ade},
  {Atad-Ettedgui}, {Berry}, {Craig}, {Gao}, {Gibb}, {Hilton}, {Hollister},
  {Kycia}, {Lunney}, {McGregor}, {Montgomery}, {Parkes}, {Tilanus}, {Ullom},
  {Walther}, {Walton}, {Woodcraft}, {Amiri}, {Atkinson}, {Burger}, {Chuter},
  {Coulson}, {Doriese}, {Dunare}, {Economou}, {Niemack}, {Parsons},
  {Reintsema}, {Sibthorpe}, {Smail}, {Sudiwala}, \&
  {Thomas}}]{2013MNRAS.430.2513H}
{Holland}, W.~S., {Bintley}, D., {Chapin}, E.~L., {et~al.} 2013, \mnras, 430,
  2513, \dodoi{10.1093/mnras/sts612}

\bibitem[{Hull {et~al.}(2020)Hull, Gouellec, Girart, Tobin, \&
  Bourke}]{Hull_2020}
Hull, C. L.~H., Gouellec, V. J. M.~L., Girart, J.~M., Tobin, J.~J., \& Bourke,
  T.~L. 2020, \apj, 892, 152, \dodoi{10.3847/1538-4357/ab5809}

\bibitem[{{Hull} {et~al.}(2017){Hull}, {Girart}, {Tychoniec}, {Rao},
  {Cort{\'e}s}, {Pokhrel}, {Zhang}, {Houde}, {Dunham}, {Kristensen}, {Lai},
  {Li}, \& {Plambeck}}]{2017ApJ...847...92H}
{Hull}, C. L.~H., {Girart}, J.~M., {Tychoniec}, {\L}., {et~al.} 2017, \apj,
  847, 92, \dodoi{10.3847/1538-4357/aa7fe9}

\bibitem[{{Hwang} {et~al.}(2023){Hwang}, {Pattle}, {Parsons}, {Go}, \&
  {Kim}}]{hwang2023}
{Hwang}, J., {Pattle}, K., {Parsons}, H., {Go}, M., \& {Kim}, J. 2023, \aj,
  165, 198, \dodoi{10.3847/1538-3881/acc460}

\bibitem[{{Jijina} {et~al.}(1999){Jijina}, {Myers}, \& {Adams}}]{jijina1999}
{Jijina}, J., {Myers}, P.~C., \& {Adams}, F.~C. 1999, \apjs, 125, 161,
  \dodoi{10.1086/313268}

\bibitem[{{Karoly} {et~al.}(2020){Karoly}, {Soam}, {Andersson}, {Coud{\'e}},
  {Bastien}, {Vaillancourt}, \& {Lee}}]{2020ApJ...900..181K}
{Karoly}, J., {Soam}, A., {Andersson}, B.~G., {et~al.} 2020, \apj, 900, 181,
  \dodoi{10.3847/1538-4357/abad37}

\bibitem[{{Kim} {et~al.}(2020){Kim}, {Tatematsu}, {Liu}, {Yi}, {He}, {Hirano},
  {Liu}, {Choi}, {Sanhueza}, {T{\'o}th}, {Evans}, {Feng}, {Juvela}, {Kim},
  {Vastel}, {Lee}, {Nguyễn Lu'o'ng}, {Kang}, {Ristorcelli}, {Feh{\'e}r},
  {Wu}, {Ohashi}, {Wang}, {Kandori}, {Hirota}, {Sakai}, {Lu}, {Thompson},
  {Fuller}, {Li}, {Shinnaga}, \& {Kim}}]{2020ApJS..249...33K}
{Kim}, G., {Tatematsu}, K., {Liu}, T., {et~al.} 2020, \apjs, 249, 33,
  \dodoi{10.3847/1538-4365/aba746}

\bibitem[{{Kirchschlager} {et~al.}(2019){Kirchschlager}, {Bertrang}, \&
  {Flock}}]{2019MNRAS.488.1211K}
{Kirchschlager}, F., {Bertrang}, G. H.~M., \& {Flock}, M. 2019, \mnras, 488,
  1211, \dodoi{10.1093/mnras/stz1763}

\bibitem[{{Kirk} {et~al.}(2018){Kirk}, {Hatchell}, {Johnstone}, {Berry},
  {Jenness}, {Buckle}, {Mairs}, {Rosolowsky}, {Di Francesco}, {Sadavoy},
  {Currie}, {Broekhoven-Fiene}, {Mottram}, {Pattle}, {Matthews}, {Knee},
  {Moriarty-Schieven}, {Duarte-Cabral}, {Tisi}, \&
  {Ward-Thompson}}]{2018ApJS..238....8K}
{Kirk}, H., {Hatchell}, J., {Johnstone}, D., {et~al.} 2018, \apjs, 238, 8,
  \dodoi{10.3847/1538-4365/aada7f}

\bibitem[{Krumholz \& Federrath(2019)}]{krumholz2019}
Krumholz, M.~R., \& Federrath, C. 2019, \frass, 6,
  \dodoi{10.3389/fspas.2019.00007}

\bibitem[{{Kwon} {et~al.}(2022){Kwon}, {Pattle}, {Sadavoy}, {Hull},
  {Johnstone}, {Ward-Thompson}, {Francesco}, {Koch}, {Furuya}, {Doi}, {Le
  Gouellec}, {Hwang}, {Lyo}, {Soam}, {Tang}, {Hoang}, {Kirchschlager},
  {Eswaraiah}, {Fanciullo}, {Kim}, {Onaka}, {K{\"o}nyves}, {Kang}, {Lee},
  {Tamura}, {Bastien}, {Hasegawa}, {Lai}, {Qiu}, {Berry}, {Arzoumanian},
  {Bourke}, {Byun}, {Chen}, {Chen}, {Chen}, {Chen}, {Ching}, {Cho}, {Choi},
  {Choi}, {Chrysostomou}, {Chung}, {Coud{\'e}}, {Dai}, {Diep}, {Duan}, {Duan},
  {Eden}, {Fiege}, {Fissel}, {Franzmann}, {Friberg}, {Friesen}, {Fuller},
  {Gledhill}, {Graves}, {Greaves}, {Griffin}, {Gu}, {Han}, {Hatchell},
  {Hayashi}, {Houde}, {Inoue}, {Inutsuka}, {Iwasaki}, {Jeong}, {Kang},
  {Karoly}, {Kataoka}, {Kawabata}, {Kemper}, {Kim}, {Kim}, {Kim}, {Kim}, {Kim},
  {Kirk}, {Kobayashi}, {Kusune}, {Kwon}, {Lacaille}, {Law}, {Lee}, {Lee},
  {Lee}, {Lee}, {Lee}, {Li}, {Li}, {Li}, {Lin}, {Liu}, {Liu}, {Liu}, {Liu},
  {Lu}, {Mairs}, {Matsumura}, {Matthews}, {Moriarty-Schieven}, {Nagata},
  {Nakamura}, {Nakanishi}, {Ngoc}, {Ohashi}, {Park}, {Parsons}, {Peretto},
  {Priestley}, {Pyo}, {Qian}, {Rao}, {Rawlings}, {Rawlings}, {Retter},
  {Richer}, {Rigby}, {Saito}, {Savini}, {Seta}, {Shimajiri}, {Shinnaga},
  {Tahani}, {Tang}, {Tomisaka}, {Tram}, {Tsukamoto}, {Viti}, {Wang}, {Wang},
  {Whitworth}, {Wu}, {Xie}, {Yen}, {Yoo}, {Yuan}, {Yun}, {Zenko}, {Zhang},
  {Zhang}, {Zhang}, {Zhou}, {Zhu}, {Looze}, {Andr{\'e}}, {Dowell}, {Eyres},
  {Falle}, {Robitaille}, \& {Loo}}]{2022ApJ...926..163K}
{Kwon}, W., {Pattle}, K., {Sadavoy}, S., {et~al.} 2022, \apj, 926, 163,
  \dodoi{10.3847/1538-4357/ac4bbe}

\bibitem[{{Lazarian} \& {Hoang}(2007)}]{2007MNRAS.378..910L}
{Lazarian}, A., \& {Hoang}, T. 2007, \mnras, 378, 910,
  \dodoi{10.1111/j.1365-2966.2007.11817.x}

\bibitem[{{Lee} \& {Ho}(2005)}]{2005ApJ...624..841L}
{Lee}, C.-F., \& {Ho}, P. T.~P. 2005, \apj, 624, 841, \dodoi{10.1086/429535}

\bibitem[{{Lee} {et~al.}(2002){Lee}, {Mundy}, {Stone}, \&
  {Ostriker}}]{2002ApJ...576..294L}
{Lee}, C.-F., {Mundy}, L.~G., {Stone}, J.~M., \& {Ostriker}, E.~C. 2002, \apj,
  576, 294, \dodoi{10.1086/341540}

\bibitem[{{Liu} {et~al.}(2022){Liu}, {Zhang}, \& {Qiu}}]{Liu2022FrASS}
{Liu}, J., {Zhang}, Q., \& {Qiu}, K. 2022, \frass, 9, 943556,
  \dodoi{10.3389/fspas.2022.943556}

\bibitem[{{Liu} {et~al.}(2019){Liu}, {Qiu}, {Berry}, {Di Francesco}, {Bastien},
  {Koch}, {Furuya}, {Kim}, {Coud{\'e}}, {Lee}, {Soam}, {Eswaraiah}, {Li},
  {Hwang}, {Lyo}, {Pattle}, {Hasegawa}, {Kwon}, {Lai}, {Ward-Thompson},
  {Ching}, {Chen}, {Gu}, {Li}, {Li}, {Liu}, {Qian}, {Wang}, {Yuan}, {Zhang},
  {Zhang}, {Zhang}, {Zhou}, {Zhu}, {Andr{\'e}}, {Arzoumanian}, {Aso}, {Byun},
  {Chen}, {Chen}, {Chen}, {Cho}, {Choi}, {Chrysostomou}, {Chung}, {Doi},
  {Drabek-Maunder}, {Dowell}, {Eyres}, {Falle}, {Fanciullo}, {Fiege},
  {Franzmann}, {Friberg}, {Friesen}, {Fuller}, {Gledhill}, {Graves}, {Greaves},
  {Griffin}, {Han}, {Hatchell}, {Hayashi}, {Hoang}, {Holland}, {Houde},
  {Inoue}, {Inutsuka}, {Iwasaki}, {Jeong}, {Johnstone}, {Kanamori}, {Kang},
  {Kang}, {Kang}, {Kataoka}, {Kawabata}, {Kemper}, {Kim}, {Kim}, {Kim}, {Kim},
  {Kim}, {Kirk}, {Kobayashi}, {Kusune}, {Kwon}, {Lacaille}, {Lee}, {Lee},
  {Lee}, {Lee}, {Liu}, {Liu}, {van Loo}, {Mairs}, {Matsumura}, {Matthews},
  {Moriarty-Schieven}, {Nagata}, {Nakamura}, {Nakanishi}, {Ohashi}, {Onaka},
  {Parker}, {Parsons}, {Pascale}, {Peretto}, {Pon}, {Pyo}, {Rao}, {Rawlings},
  {Retter}, {Richer}, {Rigby}, {Robitaille}, {Sadavoy}, {Saito}, {Savini},
  {Scaife}, {Seta}, {Shinnaga}, {Tamura}, {Tang}, {Tomisaka}, {Tsukamoto},
  {Wang}, {Whitworth}, {Yen}, {Yoo}, \& {Zenko}}]{2019ApJ...877...43L}
{Liu}, J., {Qiu}, K., {Berry}, D., {et~al.} 2019, \apj, 877, 43,
  \dodoi{10.3847/1538-4357/ab0958}

\bibitem[{{Loinard} {et~al.}(2008){Loinard}, {Torres}, {Mioduszewski}, \&
  {Rodr{\'\i}guez}}]{2008ApJ...675L..29L}
{Loinard}, L., {Torres}, R.~M., {Mioduszewski}, A.~J., \& {Rodr{\'\i}guez},
  L.~F. 2008, \apjl, 675, L29, \dodoi{10.1086/529548}

\bibitem[{{Loren}(1989)}]{loren1989}
{Loren}, R.~B. 1989, \apj, 338, 902, \dodoi{10.1086/167244}

\bibitem[{{Lyo} {et~al.}(2021){Lyo}, {Kim}, {Sadavoy}, {Johnstone}, {Berry},
  {Pattle}, {Kwon}, {Bastien}, {Onaka}, {Di Francesco}, {Kang}, {Furuya},
  {Hull}, {Tamura}, {Koch}, {Ward-Thompson}, {Hasegawa}, {Hoang},
  {Arzoumanian}, {Won Lee}, {Lee}, {Byun}, {Kirchschlager}, {Doi}, {Kim},
  {Hwang}, {Diep}, {Fanciullo}, {Lee}, {Park}, {Yoo}, {Chung}, {Whitworth},
  {Mairs}, {Soam}, {Liu}, {Tang}, {Coud{\'e}}, {Andr{\'e}}, {Bourke}, {Vivien
  Chen}, {Chen}, {Ping Chen}, {Chen}, {Ching}, {Cho}, {Choi}, {Choi},
  {Chrysostomou}, {Dai}, {Dowell}, {Duan}, {Duan}, {Eden}, {Eswaraiah},
  {Eyres}, {Fiege}, {Fissel}, {Franzmann}, {Friberg}, {Friesen}, {Fuller},
  {Gledhill}, {Graves}, {Greaves}, {Griffin}, {Gu}, {Han}, {Hatchell},
  {Hayashi}, {Houde}, {Inoue}, {Inutsuka}, {Iwasaki}, {Jeong}, {Kang},
  {Kataoka}, {Kawabata}, {Kemper}, {Kim}, {Kim}, {Kim}, {Kim}, {Kirk},
  {Kobayashi}, {K{\"o}nyves}, {Kusune}, {Kwon}, {Lacaille}, {Lai}, {Law},
  {Lee}, {Lee}, {Lee}, {Li}, {Li}, {Li}, {Liu}, {Liu}, {Liu}, {Lu},
  {Matsumura}, {Matthews}, {Moriarty-Schieven}, {Nagata}, {Nakamura},
  {Nakanishi}, {Bich Ngoc}, {Ohashi}, {Parsons}, {Peretto}, {Priestley}, {Pyo},
  {Qian}, {Qiu}, {Rao}, {Rawlings}, {Rawlings}, {Retter}, {Richer}, {Rigby},
  {Saito}, {Savini}, {Scaife}, {Seta}, {Shimajiri}, {Shinnaga}, {Tahani},
  {Tang}, {Tomisaka}, {Tram}, {Tsukamoto}, {Viti}, {Wang}, {Wang}, {Xie},
  {Yen}, {Yuan}, {Yun}, {Zenko}, {Zhang}, {Zhang}, {Zhang}, {Zhou}, {Zhu}, {de
  Looze}, {Dowell}, {Falle}, {Robitaille}, \& {van Loo}}]{2021ApJ...918...85L}
{Lyo}, A.~R., {Kim}, J., {Sadavoy}, S., {et~al.} 2021, \apj, 918, 85,
  \dodoi{10.3847/1538-4357/ac0ce9}

\bibitem[{{Mairs} {et~al.}(2015){Mairs}, {Johnstone}, {Kirk}, {Graves},
  {Buckle}, {Beaulieu}, {Berry}, {Broekhoven-Fiene}, {Currie}, {Fich},
  {Hatchell}, {Jenness}, {Mottram}, {Nutter}, {Pattle}, {Pineda}, {Salji}, {di
  Francesco}, {Hogerheijde}, {Ward-Thompson}, \& {JCMT Gould Belt survey
  Team}}]{2015MNRAS.454.2557M}
{Mairs}, S., {Johnstone}, D., {Kirk}, H., {et~al.} 2015, \mnras, 454, 2557,
  \dodoi{10.1093/mnras/stv2192}

\bibitem[{{Mairs} {et~al.}(2021){Mairs}, {Dempsey}, {Bell}, {Parsons},
  {Currie}, {Friberg}, {Jiang}, {Tetarenko}, {Bintley}, {Cookson}, {Li},
  {Rawlings}, {Wouterloot}, {Berry}, {Graves}, {Mizuno}, {Acohido}, {Clark},
  {Cox}, {Fuchs}, {Hoge}, {Kemp}, {Lee}, {Matulonis}, {Montgomerie}, {Silva},
  \& {Smith}}]{2021AJ....162..191M}
{Mairs}, S., {Dempsey}, J.~T., {Bell}, G.~S., {et~al.} 2021, \aj, 162, 191,
  \dodoi{10.3847/1538-3881/ac18bf}

\bibitem[{{Mathieu} {et~al.}(1988){Mathieu}, {Benson}, {Fuller}, {Myers}, \&
  {Schild}}]{1988ApJ...330..385M}
{Mathieu}, R.~D., {Benson}, P.~J., {Fuller}, G.~A., {Myers}, P.~C., \&
  {Schild}, R.~E. 1988, \apj, 330, 385, \dodoi{10.1086/166478}

\bibitem[{{Matthews} {et~al.}(2009){Matthews}, {McPhee}, {Fissel}, \&
  {Curran}}]{matthews09}
{Matthews}, B.~C., {McPhee}, C.~A., {Fissel}, L.~M., \& {Curran}, R.~L. 2009,
  \apjs, 182, 143, \dodoi{10.1088/0067-0049/182/1/143}

\bibitem[{{Moriarty-Schieven} {et~al.}(2006){Moriarty-Schieven}, {Johnstone},
  {Bally}, \& {Jenness}}]{2006ApJ...645..357M}
{Moriarty-Schieven}, G.~H., {Johnstone}, D., {Bally}, J., \& {Jenness}, T.
  2006, \apj, 645, 357, \dodoi{10.1086/500357}

\bibitem[{{Myers}(2017)}]{2017ApJ...838...10M}
{Myers}, P.~C. 2017, \apj, 838, 10, \dodoi{10.3847/1538-4357/aa5fa8}

\bibitem[{{Myers} {et~al.}(1988){Myers}, {Heyer}, {Snell}, \&
  {Goldsmith}}]{myers1988}
{Myers}, P.~C., {Heyer}, M., {Snell}, R.~L., \& {Goldsmith}, P.~F. 1988, \apj,
  324, 907, \dodoi{10.1086/165948}

\bibitem[{{Nakano} \& {Nakamura}(1978)}]{1978PASJ...30..671N}
{Nakano}, T., \& {Nakamura}, T. 1978, \pasj, 30, 671

\bibitem[{{Ostriker} {et~al.}(2001){Ostriker}, {Stone}, \&
  {Gammie}}]{2001ApJ...546..980O}
{Ostriker}, E.~C., {Stone}, J.~M., \& {Gammie}, C.~F. 2001, \apj, 546, 980,
  \dodoi{10.1086/318290}

\bibitem[{{Parsons} {et~al.}(2018){Parsons}, {Dempsey}, {Thomas}, {Berry},
  {Currie}, {Friberg}, {Wouterloot}, {Chrysostomou}, {Graves}, {Tilanus},
  {Bell}, \& {Rawlings}}]{harrietco18}
{Parsons}, H., {Dempsey}, J.~T., {Thomas}, H.~S., {et~al.} 2018, \apjs, 234,
  22, \dodoi{10.3847/1538-4365/aa989c}

\bibitem[{{Pattle} {et~al.}(2022{\natexlab{a}}){Pattle}, {Fissel}, {Tahani},
  {Liu}, \& {Ntormousi}}]{2022arXiv220311179P}
{Pattle}, K., {Fissel}, L., {Tahani}, M., {Liu}, T., \& {Ntormousi}, E.
  2022{\natexlab{a}}, arXiv e-prints, arXiv:2203.11179,
  \dodoi{10.48550/arXiv.2203.11179}

\bibitem[{{Pattle} {et~al.}(2015){Pattle}, {Ward-Thompson}, {Kirk}, {White},
  {Drabek-Maunder}, {Buckle}, {Beaulieu}, {Berry}, {Broekhoven-Fiene},
  {Currie}, {Fich}, {Hatchell}, {Kirk}, {Jenness}, {Johnstone}, {Mottram},
  {Nutter}, {Pineda}, {Quinn}, {Salji}, {Tisi}, {Walker-Smith}, {di Francesco},
  {Hogerheijde}, {Andr{\'e}}, {Bastien}, {Bresnahan}, {Butner}, {Chen},
  {Chrysostomou}, {Coude}, {Davis}, {Duarte-Cabral}, {Fiege}, {Friberg},
  {Friesen}, {Fuller}, {Graves}, {Greaves}, {Gregson}, {Griffin}, {Holland},
  {Joncas}, {Knee}, {K{\"o}nyves}, {Mairs}, {Marsh}, {Matthews},
  {Moriarty-Schieven}, {Rawlings}, {Richer}, {Robertson}, {Rosolowsky},
  {Rumble}, {Sadavoy}, {Spinoglio}, {Thomas}, {Tothill}, {Viti}, {Wouterloot},
  {Yates}, \& {Zhu}}]{2015MNRAS.450.1094P}
{Pattle}, K., {Ward-Thompson}, D., {Kirk}, J.~M., {et~al.} 2015, \mnras, 450,
  1094, \dodoi{10.1093/mnras/stv376}

\bibitem[{{Pattle} {et~al.}(2017){Pattle}, {Ward-Thompson}, {Berry},
  {Hatchell}, {Chen}, {Pon}, {Koch}, {Kwon}, {Kim}, {Bastien}, {Cho},
  {Coud{\'e}}, {Di Francesco}, {Fuller}, {Furuya}, {Graves}, {Johnstone},
  {Kirk}, {Kwon}, {Lee}, {Matthews}, {Mottram}, {Parsons}, {Sadavoy},
  {Shinnaga}, {Soam}, {Hasegawa}, {Lai}, {Qiu}, \&
  {Friberg}}]{2017ApJ...846..122P}
{Pattle}, K., {Ward-Thompson}, D., {Berry}, D., {et~al.} 2017, \apj, 846, 122,
  \dodoi{10.3847/1538-4357/aa80e5}

\bibitem[{Pattle {et~al.}(2019)Pattle, Lai, Hasegawa, Wang, Furuya,
  Ward-Thompson, Bastien, Coud{\'{e}}, Eswaraiah, Fanciullo, di~Francesco,
  Hoang, Kim, Kwon, Lee, Liu, Liu, Matsumura, Onaka, Sadavoy, \&
  Soam}]{Pattle_2019}
Pattle, K., Lai, S.-P., Hasegawa, T., {et~al.} 2019, \apj, 880, 27,
  \dodoi{10.3847/1538-4357/ab286f}

\bibitem[{{Pattle} {et~al.}(2021){Pattle}, {Lai}, {Di Francesco}, {Sadavoy},
  {Ward-Thompson}, {Johnstone}, {Hoang}, {Arzoumanian}, {Bastien}, {Bourke},
  {Coud{\'e}}, {Doi}, {Eswaraiah}, {Fanciullo}, {Furuya}, {Hwang}, {Hull},
  {Kang}, {Kim}, {Kirchschlager}, {Kwon}, {Kwon}, {Lee}, {Liu}, {Redman},
  {Soam}, {Tahani}, {Tamura}, \& {Tang}}]{2021ApJ...907...88P}
{Pattle}, K., {Lai}, S.-P., {Di Francesco}, J., {et~al.} 2021, \apj, 907, 88,
  \dodoi{10.3847/1538-4357/abcc6c}

\bibitem[{{Pattle} {et~al.}(2022{\natexlab{b}}){Pattle}, {Lai}, {Sadavoy},
  {Coud{\'e}}, {Wolf}, {Furuya}, {Kwon}, {Lee}, \&
  {Zielinski}}]{2022MNRAS.515.1026P}
{Pattle}, K., {Lai}, S.-P., {Sadavoy}, S., {et~al.} 2022{\natexlab{b}}, \mnras,
  515, 1026, \dodoi{10.1093/mnras/stac1356}

\bibitem[{{Planck Collaboration} {et~al.}(2016{\natexlab{a}}){Planck
  Collaboration}, {Ade}, {Aghanim}, {Alves}, {Arnaud}, {Arzoumanian},
  {Ashdown}, {Aumont}, {Baccigalupi}, {Banday}, {Barreiro}, {Bartolo},
  {Battaner}, {Benabed}, {Beno{\^\i}t}, {Benoit-L{\'e}vy}, {Bernard},
  {Bersanelli}, {Bielewicz}, {Bock}, {Bonavera}, {Bond}, {Borrill}, {Bouchet},
  {Boulanger}, {Bracco}, {Burigana}, {Calabrese}, {Cardoso}, {Catalano},
  {Chiang}, {Christensen}, {Colombo}, {Combet}, {Couchot}, {Crill}, {Curto},
  {Cuttaia}, {Danese}, {Davies}, {Davis}, {de Bernardis}, {de Rosa}, {de
  Zotti}, {Delabrouille}, {Dickinson}, {Diego}, {Dole}, {Donzelli}, {Dor{\'e}},
  {Douspis}, {Ducout}, {Dupac}, {Efstathiou}, {Elsner}, {En{\ss}lin},
  {Eriksen}, {Falceta-Gon{\c{c}}alves}, {Falgarone}, {Ferri{\`e}re}, {Finelli},
  {Forni}, {Frailis}, {Fraisse}, {Franceschi}, {Frejsel}, {Galeotta}, {Galli},
  {Ganga}, {Ghosh}, {Giard}, {Gjerl{\o}w}, {Gonz{\'a}lez-Nuevo}, {G{\'o}rski},
  {Gregorio}, {Gruppuso}, {Gudmundsson}, {Guillet}, {Harrison}, {Helou},
  {Hennebelle}, {Henrot-Versill{\'e}}, {Hern{\'a}ndez-Monteagudo}, {Herranz},
  {Hildebrandt}, {Hivon}, {Holmes}, {Hornstrup}, {Huffenberger}, {Hurier},
  {Jaffe}, {Jaffe}, {Jones}, {Juvela}, {Keih{\"a}nen}, {Keskitalo}, {Kisner},
  {Knoche}, {Kunz}, {Kurki-Suonio}, {Lagache}, {Lamarre}, {Lasenby},
  {Lattanzi}, {Lawrence}, {Leonardi}, {Levrier}, {Liguori}, {Lilje},
  {Linden-V{\o}rnle}, {L{\'o}pez-Caniego}, {Lubin}, {Mac{\'\i}as-P{\'e}rez},
  {Maino}, {Mandolesi}, {Mangilli}, {Maris}, {Martin},
  {Mart{\'\i}nez-Gonz{\'a}lez}, {Masi}, {Matarrese}, {Melchiorri}, {Mendes},
  {Mennella}, {Migliaccio}, {Miville-Desch{\^e}nes}, {Moneti}, {Montier},
  {Morgante}, {Mortlock}, {Munshi}, {Murphy}, {Naselsky}, {Nati},
  {Netterfield}, {Noviello}, {Novikov}, {Novikov}, {Oppermann}, {Oxborrow},
  {Pagano}, {Pajot}, {Paladini}, {Paoletti}, {Pasian}, {Perotto}, {Pettorino},
  {Piacentini}, {Piat}, {Pierpaoli}, {Pietrobon}, {Plaszczynski},
  {Pointecouteau}, {Polenta}, {Ponthieu}, {Pratt}, {Prunet}, {Puget}, {Rachen},
  {Reinecke}, {Remazeilles}, {Renault}, {Renzi}, {Ristorcelli}, {Rocha},
  {Rossetti}, {Roudier}, {Rubi{\~n}o-Mart{\'\i}n}, {Rusholme}, {Sandri},
  {Santos}, {Savelainen}, {Savini}, {Scott}, {Soler}, {Stolyarov}, {Sudiwala},
  {Sutton}, {Suur-Uski}, {Sygnet}, {Tauber}, {Terenzi}, {Toffolatti}, {Tomasi},
  {Tristram}, {Tucci}, {Umana}, {Valenziano}, {Valiviita}, {Van Tent},
  {Vielva}, {Villa}, {Wade}, {Wandelt}, {Wehus}, {Ysard}, {Yvon}, \&
  {Zonca}}]{2016A&A...586A.138P}
{Planck Collaboration}, {Ade}, P.~A.~R., {Aghanim}, N., {et~al.}
  2016{\natexlab{a}}, \aap, 586, A138, \dodoi{10.1051/0004-6361/201525896}

\bibitem[{{Planck Collaboration} {et~al.}(2016{\natexlab{b}}){Planck
  Collaboration}, {Ade}, {Aghanim}, {Alves}, {Aniano}, {Arnaud}, {Ashdown},
  {Aumont}, {Baccigalupi}, {Banday}, {Barreiro}, {Bartolo}, {Battaner},
  {Benabed}, {Benoit-L{\'e}vy}, {Bernard}, {Bersanelli}, {Bielewicz},
  {Bonaldi}, {Bonavera}, {Bond}, {Borrill}, {Bouchet}, {Boulanger}, {Burigana},
  {Butler}, {Calabrese}, {Cardoso}, {Catalano}, {Chamballu}, {Chiang},
  {Christensen}, {Clements}, {Colombi}, {Colombo}, {Couchot}, {Crill}, {Curto},
  {Cuttaia}, {Danese}, {Davies}, {Davis}, {de Bernardis}, {de Rosa}, {de
  Zotti}, {Delabrouille}, {Dickinson}, {Diego}, {Dole}, {Donzelli}, {Dor{\'e}},
  {Douspis}, {Draine}, {Ducout}, {Dupac}, {Efstathiou}, {Elsner}, {En{\ss}lin},
  {Eriksen}, {Falgarone}, {Finelli}, {Forni}, {Frailis}, {Fraisse},
  {Franceschi}, {Frejsel}, {Galeotta}, {Galli}, {Ganga}, {Ghosh}, {Giard},
  {Gjerl{\o}w}, {Gonz{\'a}lez-Nuevo}, {G{\'o}rski}, {Gregorio}, {Gruppuso},
  {Guillet}, {Hansen}, {Hanson}, {Harrison}, {Henrot-Versill{\'e}},
  {Hern{\'a}ndez-Monteagudo}, {Herranz}, {Hildebrandt}, {Hivon}, {Holmes},
  {Hovest}, {Huffenberger}, {Hurier}, {Jaffe}, {Jaffe}, {Jones},
  {Keih{\"a}nen}, {Keskitalo}, {Kisner}, {Kneissl}, {Knoche}, {Kunz},
  {Kurki-Suonio}, {Lagache}, {Lamarre}, {Lasenby}, {Lattanzi}, {Lawrence},
  {Leonardi}, {Levrier}, {Liguori}, {Lilje}, {Linden-V{\o}rnle},
  {L{\'o}pez-Caniego}, {Lubin}, {Mac{\'\i}as-P{\'e}rez}, {Maffei}, {Maino},
  {Mandolesi}, {Maris}, {Marshall}, {Martin}, {Mart{\'\i}nez-Gonz{\'a}lez},
  {Masi}, {Matarrese}, {Mazzotta}, {Melchiorri}, {Mendes}, {Mennella},
  {Migliaccio}, {Miville-Desch{\^e}nes}, {Moneti}, {Montier}, {Morgante},
  {Mortlock}, {Munshi}, {Murphy}, {Naselsky}, {Natoli}, {N{\o}rgaard-Nielsen},
  {Novikov}, {Novikov}, {Oxborrow}, {Pagano}, {Pajot}, {Paladini}, {Paoletti},
  {Pasian}, {Perdereau}, {Perotto}, {Perrotta}, {Pettorino}, {Piacentini},
  {Piat}, {Plaszczynski}, {Pointecouteau}, {Polenta}, {Ponthieu}, {Popa},
  {Pratt}, {Prunet}, {Puget}, {Rachen}, {Reach}, {Rebolo}, {Reinecke},
  {Remazeilles}, {Renault}, {Ristorcelli}, {Rocha}, {Roudier},
  {Rubi{\~n}o-Mart{\'\i}n}, {Rusholme}, {Sandri}, {Santos}, {Scott}, {Spencer},
  {Stolyarov}, {Sudiwala}, {Sunyaev}, {Sutton}, {Suur-Uski}, {Sygnet},
  {Tauber}, {Terenzi}, {Toffolatti}, {Tomasi}, {Tristram}, {Tucci}, {Umana},
  {Valenziano}, {Valiviita}, {Van Tent}, {Vielva}, {Villa}, {Wade}, {Wandelt},
  {Wehus}, {Ysard}, {Yvon}, {Zacchei}, \& {Zonca}}]{2016A&A...586A.132P}
---. 2016{\natexlab{b}}, \aap, 586, A132, \dodoi{10.1051/0004-6361/201424945}

\bibitem[{{Planck Collaboration} {et~al.}(2016{\natexlab{c}}){Planck
  Collaboration}, {Ade}, {Aghanim}, {Arnaud}, {Ashdown}, {Aumont},
  {Baccigalupi}, {Banday}, {Barreiro}, {Bartolo}, {Battaner}, {Benabed},
  {Beno{\^\i}t}, {Benoit-L{\'e}vy}, {Bernard}, {Bersanelli}, {Bielewicz},
  {Bonaldi}, {Bonavera}, {Bond}, {Borrill}, {Bouchet}, {Boulanger}, {Bucher},
  {Burigana}, {Butler}, {Calabrese}, {Catalano}, {Chamballu}, {Chiang},
  {Christensen}, {Clements}, {Colombi}, {Colombo}, {Combet}, {Couchot},
  {Coulais}, {Crill}, {Curto}, {Cuttaia}, {Danese}, {Davies}, {Davis}, {de
  Bernardis}, {de Rosa}, {de Zotti}, {Delabrouille}, {D{\'e}sert}, {Dickinson},
  {Diego}, {Dole}, {Donzelli}, {Dor{\'e}}, {Douspis}, {Ducout}, {Dupac},
  {Efstathiou}, {Elsner}, {En{\ss}lin}, {Eriksen}, {Falgarone}, {Fergusson},
  {Finelli}, {Forni}, {Frailis}, {Fraisse}, {Franceschi}, {Frejsel},
  {Galeotta}, {Galli}, {Ganga}, {Giard}, {Giraud-H{\'e}raud}, {Gjerl{\o}w},
  {Gonz{\'a}lez-Nuevo}, {G{\'o}rski}, {Gratton}, {Gregorio}, {Gruppuso},
  {Gudmundsson}, {Hansen}, {Hanson}, {Harrison}, {Helou},
  {Henrot-Versill{\'e}}, {Hern{\'a}ndez-Monteagudo}, {Herranz}, {Hildebrandt},
  {Hivon}, {Hobson}, {Holmes}, {Hornstrup}, {Hovest}, {Huffenberger}, {Hurier},
  {Jaffe}, {Jaffe}, {Jones}, {Juvela}, {Keih{\"a}nen}, {Keskitalo}, {Kisner},
  {Knoche}, {Kunz}, {Kurki-Suonio}, {Lagache}, {Lamarre}, {Lasenby},
  {Lattanzi}, {Lawrence}, {Leonardi}, {Lesgourgues}, {Levrier}, {Liguori},
  {Lilje}, {Linden-V{\o}rnle}, {L{\'o}pez-Caniego}, {Lubin},
  {Mac{\'\i}as-P{\'e}rez}, {Maggio}, {Maino}, {Mandolesi}, {Mangilli},
  {Marshall}, {Martin}, {Mart{\'\i}nez-Gonz{\'a}lez}, {Masi}, {Matarrese},
  {Mazzotta}, {McGehee}, {Melchiorri}, {Mendes}, {Mennella}, {Migliaccio},
  {Mitra}, {Miville-Desch{\^e}nes}, {Moneti}, {Montier}, {Morgante},
  {Mortlock}, {Moss}, {Munshi}, {Murphy}, {Naselsky}, {Nati}, {Natoli},
  {Netterfield}, {N{\o}rgaard-Nielsen}, {Noviello}, {Novikov}, {Novikov},
  {Oxborrow}, {Paci}, {Pagano}, {Pajot}, {Paladini}, {Paoletti}, {Pasian},
  {Patanchon}, {Pearson}, {Pelkonen}, {Perdereau}, {Perotto}, {Perrotta},
  {Pettorino}, {Piacentini}, {Piat}, {Pierpaoli}, {Pietrobon}, {Plaszczynski},
  {Pointecouteau}, {Polenta}, {Pratt}, {Pr{\'e}zeau}, {Prunet}, {Puget},
  {Rachen}, {Reach}, {Rebolo}, {Reinecke}, {Remazeilles}, {Renault}, {Renzi},
  {Ristorcelli}, {Rocha}, {Rosset}, {Rossetti}, {Roudier},
  {Rubi{\~n}o-Mart{\'\i}n}, {Rusholme}, {Sandri}, {Santos}, {Savelainen},
  {Savini}, {Scott}, {Seiffert}, {Shellard}, {Spencer}, {Stolyarov},
  {Sudiwala}, {Sunyaev}, {Sutton}, {Suur-Uski}, {Sygnet}, {Tauber}, {Terenzi},
  {Toffolatti}, {Tomasi}, {Tristram}, {Tucci}, {Tuovinen}, {Umana},
  {Valenziano}, {Valiviita}, {Van Tent}, {Vielva}, {Villa}, {Wade}, {Wandelt},
  {Wehus}, {Yvon}, {Zacchei}, \& {Zonca}}]{2016A&A...594A..28P}
---. 2016{\natexlab{c}}, \aap, 594, A28, \dodoi{10.1051/0004-6361/201525819}

\bibitem[{{Pontoppidan} {et~al.}(2010){Pontoppidan}, {Salyk}, {Blake}, \&
  {K{\"a}ufl}}]{pont2010}
{Pontoppidan}, K.~M., {Salyk}, C., {Blake}, G.~A., \& {K{\"a}ufl}, H.~U. 2010,
  \apjl, 722, L173, \dodoi{10.1088/2041-8205/722/2/L173}

\bibitem[{{Roy, A.} {et~al.}(2014){Roy, A.}, {Andr\'e, Ph.}, {Palmeirim, P.},
  {Attard, M.}, {K\"onyves, V.}, {Schneider, N.}, {Peretto, N.},
  {Men\'{}shchikov, A.}, {Ward-Thompson, D.}, {Kirk, J.}, {Griffin, M.},
  {Marsh, K.}, {Abergel, A.}, {Arzoumanian, D.}, {Benedettini, M.}, {Hill, T.},
  {Motte, F.}, {Nguyen Luong, Q.}, {Pezzuto, S.}, {Rivera-Ingraham, A.},
  {Roussel, H.}, {Rygl, K. L. J.}, {Spinoglio, L.}, {Stamatellos, D.}, \&
  {White, G.}}]{roy2014}
{Roy, A.}, {Andr\'e, Ph.}, {Palmeirim, P.}, {et~al.} 2014, A\&A, 562, A138,
  \dodoi{10.1051/0004-6361/201322236}

\bibitem[{{Sadavoy} {et~al.}(2013){Sadavoy}, {Di Francesco}, {Johnstone},
  {Currie}, {Drabek}, {Hatchell}, {Nutter}, {Andr{\'e}}, {Arzoumanian},
  {Benedettini}, {Bernard}, {Duarte-Cabral}, {Fallscheer}, {Friesen},
  {Greaves}, {Hennemann}, {Hill}, {Jenness}, {K{\"o}nyves}, {Matthews},
  {Mottram}, {Pezzuto}, {Roy}, {Rygl}, {Schneider-Bontemps}, {Spinoglio},
  {Testi}, {Tothill}, {Ward-Thompson}, {White}, {JCMT}, \& {Herschel Gould Belt
  Survey Teams}}]{2013ApJ...767..126S}
{Sadavoy}, S.~I., {Di Francesco}, J., {Johnstone}, D., {et~al.} 2013, \apj,
  767, 126, \dodoi{10.1088/0004-637X/767/2/126}

\bibitem[{{Schnee} {et~al.}(2010){Schnee}, {Enoch}, {Noriega-Crespo}, {Sayers},
  {Terebey}, {Caselli}, {Foster}, {Goodman}, {Kauffmann}, {Padgett}, {Rebull},
  {Sargent}, \& {Shetty}}]{2010ApJ...708..127S}
{Schnee}, S., {Enoch}, M., {Noriega-Crespo}, A., {et~al.} 2010, \apj, 708, 127,
  \dodoi{10.1088/0004-637X/708/1/127}

\bibitem[{{Shirley} {et~al.}(2005){Shirley}, {Nordhaus}, {Grcevich}, {Evans},
  {Rawlings}, \& {Tatematsu}}]{2005ApJ...632..982S}
{Shirley}, Y.~L., {Nordhaus}, M.~K., {Grcevich}, J.~M., {et~al.} 2005, \apj,
  632, 982, \dodoi{10.1086/431963}

\bibitem[{Soam {et~al.}(2018)Soam, Pattle, Ward-Thompson, Lee, Sadavoy, Koch,
  Kim, Kwon, Kwon, Arzoumanian, Berry, Hoang, Tamura, Lee, Liu, Kim, Johnstone,
  Nakamura, Lyo, Onaka, Kim, Furuya, Hasegawa, Lai, Bastien, Chung, Kim,
  Parsons, Rawlings, Mairs, Graves, Robitaille, Liu, Whitworth, Eswaraiah, Rao,
  Yoo, Houde, hyun Kang, Doi, Choi, Kang, Coud{\'{e}}, bai Li, Matsumura,
  Matthews, Pon, Francesco, Hayashi, Kawabata, ichiro Inutsuka, Qiu, Franzmann,
  Friberg, Greaves, Kirk, Li, Shinnaga, van Loo, Aso, Byun, Chen, Chen, Chen,
  Ching, Cho, Chrysostomou, Drabek-Maunder, Eyres, Fiege, Friesen, Fuller,
  Gledhill, Griffin, Gu, Hatchell, Holland, Inoue, Iwasaki, Jeong, ju~Kang,
  Kemper, Kim, Kim, Lacaille, Lee, Li, Liu, Liu, Moriarty-Schieven, Nakanishi,
  Ohashi, Peretto, Pyo, Qian, Retter, Richer, Rigby, Savini, Scaife, Tang,
  Tomisaka, Wang, Wang, Yen, Yuan, Zhang, Zhang, Zhou, Zhu, Andr{\'{e}},
  Dowell, Falle, Tsukamoto, Kanamori, Kataoka, Kobayashi, Nagata, Saito, Seta,
  Hwang, Han, Lee, \& Zenko}]{Soam_2018}
Soam, A., Pattle, K., Ward-Thompson, D., {et~al.} 2018, \apj, 861, 65,
  \dodoi{10.3847/1538-4357/aac4a6}

\bibitem[{Soler {et~al.}(2013)Soler, Hennebelle, Martin,
  Miville-Desch{\^{e}}nes, Netterfield, \& Fissel}]{Soler_2013}
Soler, J.~D., Hennebelle, P., Martin, P.~G., {et~al.} 2013, \apj, 774, 128,
  \dodoi{10.1088/0004-637x/774/2/128}

\bibitem[{{Ward-Thompson} {et~al.}(2000){Ward-Thompson}, {Kirk}, {Crutcher},
  {Greaves}, {Holland}, \& {Andr{\'e}}}]{2000ApJ...537L.135W}
{Ward-Thompson}, D., {Kirk}, J.~M., {Crutcher}, R.~M., {et~al.} 2000, \apjl,
  537, L135, \dodoi{10.1086/312764}

\bibitem[{{Ward-Thompson} \& {Whitworth}(2011)}]{2011isf..book.....W}
{Ward-Thompson}, D., \& {Whitworth}, A.~P. 2011, {An Introduction to Star
  Formation} (Cambridge University Press)

\bibitem[{{Ward-Thompson} {et~al.}(2007){Ward-Thompson}, {Di Francesco},
  {Hatchell}, {Hogerheijde}, {Nutter}, {Bastien}, {Basu}, {Bonnell}, {Bowey},
  {Brunt}, {Buckle}, {Butner}, {Cavanagh}, {Chrysostomou}, {Curtis}, {Davis},
  {Dent}, {van Dishoeck}, {Edmunds}, {Fich}, {Fiege}, {Fissel}, {Friberg},
  {Friesen}, {Frieswijk}, {Fuller}, {Gosling}, {Graves}, {Greaves}, {Helmich},
  {Hills}, {Holland}, {Houde}, {Jayawardhana}, {Johnstone}, {Joncas}, {Kirk},
  {Kirk}, {Knee}, {Matthews}, {Matthews}, {Matzner}, {Moriarty-Schieven},
  {Naylor}, {Padman}, {Plume}, {Rawlings}, {Redman}, {Reid}, {Richer},
  {Shipman}, {Simpson}, {Spaans}, {Stamatellos}, {Tsamis}, {Viti}, {Weferling},
  {White}, {Whitworth}, {Wouterloot}, {Yates}, \& {Zhu}}]{2007PASP..119..855W}
{Ward-Thompson}, D., {Di Francesco}, J., {Hatchell}, J., {et~al.} 2007, \pasp,
  119, 855, \dodoi{10.1086/52127710.48550/arXiv.0707.0169}

\bibitem[{Ward-Thompson {et~al.}(2017)Ward-Thompson, Pattle, Bastien, Furuya,
  Kwon, Lai, Qiu, Berry, Choi, Coud{\'{e}}, Francesco, Hoang, Franzmann,
  Friberg, Graves, Greaves, Houde, Johnstone, Kirk, Koch, Kwon, Lee, Li,
  Matthews, Mottram, Parsons, Pon, Rao, Rawlings, Shinnaga, Sadavoy, van Loo,
  Aso, Byun, Eswaraiah, Chen, Chen, Chen, Ching, Cho, Chrysostomou, Chung, Doi,
  Drabek-Maunder, Eyres, Fiege, Friesen, Fuller, Gledhill, Griffin, Gu,
  Hasegawa, Hatchell, Hayashi, Holland, Inoue, ichiro Inutsuka, Iwasaki, Jeong,
  hyun Kang, Kang, ju~Kang, Kawabata, Kemper, Kim, Kim, Kim, Kim, Kim, Kim,
  Lacaille, Lee, Lee, Li, bai Li, Liu, Liu, Liu, Liu, Lyo, Mairs, Matsumura,
  Moriarty-Schieven, Nakamura, Nakanishi, Ohashi, Onaka, Peretto, Pyo, Qian,
  Retter, Richer, Rigby, Robitaille, Savini, Scaife, Soam, Tamura, Tang,
  Tomisaka, Wang, Wang, Whitworth, Yen, Yoo, Yuan, Zhang, Zhang, Zhou, Zhu,
  Andr{\'{e}}, Dowell, Falle, \& Tsukamoto}]{Ward_Thompson_2017}
Ward-Thompson, D., Pattle, K., Bastien, P., {et~al.} 2017, \apj, 842, 66,
  \dodoi{10.3847/1538-4357/aa70a0}

\bibitem[{{Ward-Thompson} {et~al.}(2023){Ward-Thompson}, {Karoly}, {Pattle},
  {Whitworth}, {Kirk}, {Berry}, {Bastien}, {Ching}, {Coud{\'e}}, {Hwang},
  {Kwon}, {Soam}, {Wang}, {Hasegawa}, {Lai}, {Qiu}, {Arzoumanian}, {Bourke},
  {Byun}, {Chen}, {Chen}, {Chen}, {Chen}, {Cho}, {Choi}, {Choi}, {Choi},
  {Chrysostomou}, {Chung}, {Dai}, {Debattista}, {Di Francesco}, {Diep}, {Doi},
  {Duan}, {Duan}, {Eswaraiah}, {Fanciullo}, {Fiege}, {Fissel}, {Franzmann},
  {Friberg}, {Friesen}, {Fuller}, {Furuya}, {Gledhill}, {Graves}, {Greaves},
  {Griffin}, {Gu}, {Han}, {Hayashi}, {Hoang}, {Houde}, {Hull}, {Inoue},
  {Inutsuka}, {Iwasaki}, {Jeong}, {Johnstone}, {K{\"o}nyves}, {Kang}, {Kang},
  {Kataoka}, {Kawabata}, {Kemper}, {Kim}, {Kim}, {Kim}, {Kim}, {Kim}, {Kim},
  {Kim}, {Kirchschlager}, {Kobayashi}, {Koch}, {Kusune}, {Kwon}, {Lacaille},
  {Law}, {Lee}, {Lee}, {Lee}, {Lee}, {Lee}, {Lee}, {Li}, {Li}, {Li}, {Li},
  {Lin}, {Liu}, {Liu}, {Liu}, {Liu}, {Longmore}, {Lu}, {Lyo}, {Mairs},
  {Matsumura}, {Matthews}, {Moriarty-Schieven}, {Nagata}, {Nakamura},
  {Nakanishi}, {Ngoc}, {Ohashi}, {Onaka}, {Park}, {Parsons}, {Peretto},
  {Priestley}, {Pyo}, {Qian}, {Rao}, {Rawlings}, {Rawlings}, {Retter},
  {Richer}, {Rigby}, {Sadavoy}, {Saito}, {Savini}, {Seta}, {Shimajiri},
  {Shinnaga}, {Tahani}, {Tamura}, {Tang}, {Tang}, {Tomisaka}, {Tram},
  {Tsukamoto}, {Viti}, {Wang}, {Wu}, {Xie}, {Yang}, {Yen}, {Yoo}, {Yuan},
  {Yun}, {Zenko}, {Zhang}, {Zhang}, {Zhang}, {Zhou}, {Zhu}, {de Looze},
  {Andr{\'e}}, {Dowell}, {Eden}, {Eyres}, {Falle}, {Le Gouellec}, {Poidevin},
  {Robitaille}, \& {van Loo}}]{derekl1495}
{Ward-Thompson}, D., {Karoly}, J., {Pattle}, K., {et~al.} 2023, \apj, 946, 62,
  \dodoi{10.3847/1538-4357/acbea4}

\bibitem[{{Wardle} \& {Kronberg}(1974)}]{1974ApJ...194..249W}
{Wardle}, J.~F.~C., \& {Kronberg}, P.~P. 1974, \apj, 194, 249,
  \dodoi{10.1086/153240}

\bibitem[{{Weintraub} {et~al.}(1994){Weintraub}, {Tegler}, {Kastner}, \&
  {Rettig}}]{1994ApJ...423..674W}
{Weintraub}, D.~A., {Tegler}, S.~C., {Kastner}, J.~H., \& {Rettig}, T. 1994,
  \apj, 423, 674, \dodoi{10.1086/173846}

\bibitem[{{Yang} {et~al.}(2018){Yang}, {Green}, {Evans}, {Lee}, {J{\o}rgensen},
  {Kristensen}, {Mottram}, {Herczeg}, {Karska}, {Dionatos}, {Bergin},
  {Bouwman}, {van Dishoeck}, {van Kempen}, {Larson}, \&
  {Y{\i}ld{\i}z}}]{2018ApJ...860..174Y}
{Yang}, Y.-L., {Green}, J.~D., {Evans}, Neal~J., I., {et~al.} 2018, \apj, 860,
  174, \dodoi{10.3847/1538-4357/aac2c6}

\bibitem[{{Yen} {et~al.}(2021){Yen}, {Koch}, {Hull}, {Ward-Thompson},
  {Bastien}, {Hasegawa}, {Kwon}, {Lai}, {Qiu}, {Ching}, {Chung}, {Coud{\'e}},
  {Di Francesco}, {Diep}, {Doi}, {Eswaraiah}, {Falle}, {Fuller}, {Furuya},
  {Han}, {Hatchell}, {Houde}, {Inutsuka}, {Johnstone}, {Kang}, {Kang}, {Kim},
  {Kirchschlager}, {Kwon}, {Lee}, {Lee}, {Liu}, {Liu}, {Lyo}, {Ohashi},
  {Onaka}, {Pattle}, {Sadavoy}, {Saito}, {Shinnaga}, {Soam}, {Tahani},
  {Tamura}, {Tang}, {Tang}, \& {Zhang}}]{2021ApJ...907...33Y}
{Yen}, H.-W., {Koch}, P.~M., {Hull}, C. L.~H., {et~al.} 2021, \apj, 907, 33,
  \dodoi{10.3847/1538-4357/abca99}

\bibitem[{{Yoon} {et~al.}(2021){Yoon}, {Lee}, {Lee}, {Herczeg}, {Park}, {Mace},
  {Lee}, \& {Jaffe}}]{2021ApJ...919..116Y}
{Yoon}, S.-Y., {Lee}, J.-E., {Lee}, S., {et~al.} 2021, \apj, 919, 116,
  \dodoi{10.3847/1538-4357/ac1358}

\bibitem[{{Young} {et~al.}(2006){Young}, {Bourke}, {Young}, {Evans},
  {J{\o}rgensen}, {Shirley}, {van Dishoeck}, \& {Hogerheijde}}]{young2006}
{Young}, C.~H., {Bourke}, T.~L., {Young}, K.~E., {et~al.} 2006, \aj, 132, 1998,
  \dodoi{10.1086/507334}

\end{thebibliography}
\bibliographystyle{aasjournal}

\appendix
\counterwithin{figure}{section}

\section{8$\arcsec$ Stokes \textit{I} maps}
\label{app:pix8}

\begin{figure*}
    \centering
    \includegraphics[scale=0.35,angle=0]{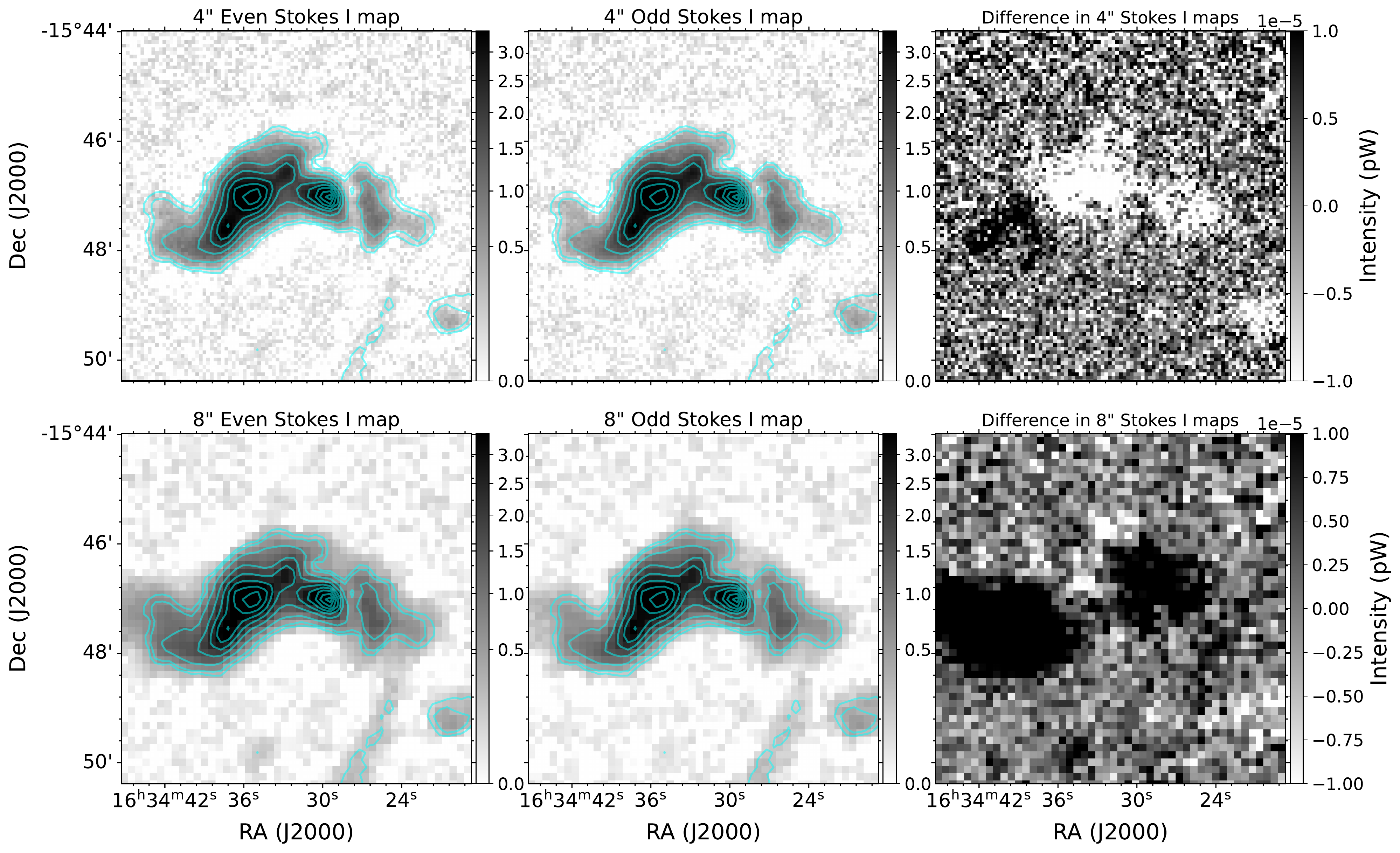}
    \caption{Results of the Jackknife Test for the data reduction technique using the auto-generated 8$\arcsec$ masks when reducing with 8$\arcsec$ pixels. Top row shows the 4$\arcsec$ Stokes \textit{I} maps from the even and odd groups as well as the difference between the groups. The color scales on the even and odd maps are $\times$10$^{-4}$\,pW.}
    \label{fig:jack8masks}
\end{figure*}

Our initial concern over using a reduction that made use of auto-generated masks from an initial 8$\arcsec$ Stokes \textit{I} map was that the emission was much more extended than in the 4$\arcsec$ maps. We do not expect to recover much large-scale flux due to inherent limitations with SCUBA-2/POL-2 and observing through the atmosphere. SCUBA-2 is fundamentally unable to measure flux on size scales larger than the array size due to the need to distinguish between atmospheric and astrophysical signal \citep{2013MNRAS.430.2513H,2013MNRAS.430.2545C}, and POL-2 is even more restricted due to its small map size and slow mapping speed \citep{2016SPIE.9914E..03F}. There are detailed discussions of SCUBA-2 large-scale flux loss compared to Herschel in \citet{2013ApJ...767..126S} and \citet{2015MNRAS.450.1094P}, and detailed discussion of the role of masking in SCUBA-2 data reduction in \citet{2015MNRAS.454.2557M} and \citet{2018ApJS..238....8K}. The first two figures in the bottom panel of Figure\,\ref{fig:jack8masks} demonstrates the additional large-scale flux we see when using auto-generated 8$\arcsec$ masks, where the contours from the Stokes \textit{I} continuum created using the regridded 4$\arcsec$ masks are plotted over the Stokes \textit{I} continuum resulting from reducing the data using the auto-generated 8$\arcsec$ masks. The contour levels are the same as those shown in Figure\,\ref{fig:cdtemp}. The background map for both even and odd groups clearly shows emission beyond the extent of the drawn contours.

However, it can also be seen that this extended emission that we see from the contours aligns well with the SPIRE 250\,$\mu$m image as mentioned in Section\,\ref{subsec:pix8}. It can be seen from Figure\,\ref{fig:herschel} that the 250\,$\mu$m dust emission extends further to the east from where the 850\,$\mu$m contours end and follows the same shape that we see from the 8$\arcsec$ emission in Figure\,\ref{fig:jack8masks}. We also see this dust morphology in all the \textit{Herschel} bands. However, it does not follow that the POL-2 Stokes \textit{I} data in these regions is well-characterized because \textit{Herschel} is able to observe extended structure.

As mentioned in Section\,\ref{subsec:pix8}, to further investigate this, we performed a Jackknife Test on the data, We divided the 26 observations into two groups, which we have designated as ‘even’ and ‘odd’. We divided the observations by just alternating between ‘even’ and ‘odd’ when the observations were ordered by date of observation. In each group, we then reduced the observations using the normal method described in Sec.\ref{subsec:data} with both 4$\arcsec$ pixels and with 8$\arcsec$ pixels. Then for each group we reduced the observations using the new method (see Sec.~\ref{subsec:pix8}), where we regridded the masks from the 4$\arcsec$ reduction in each group to 8$\arcsec$ and used those masks when running an 8$\arcsec$ reduction instead of using the auto-generated 8$\arcsec$ masks.

\begin{figure*}
    \centering
    \includegraphics[scale=0.35,angle=0]{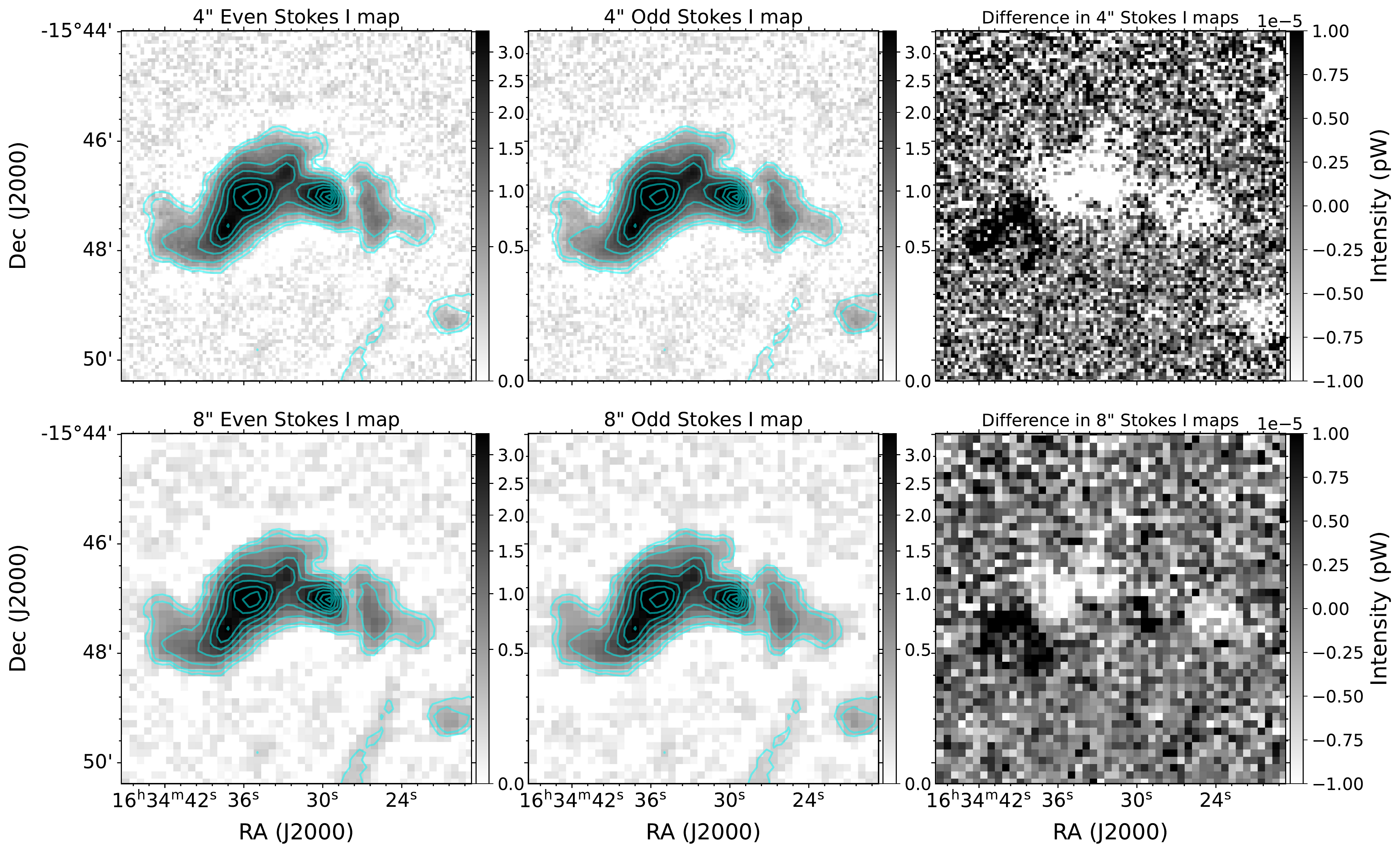}
    \caption{Results of the Jackknife Test for the data reduction technique presented in Section\,\ref{subsec:pix8}. Top row shows the 4$\arcsec$ Stokes \textit{I} maps from the even and odd groups as well as the difference between the groups. Bottom row are the 8$\arcsec$ Stokes \textit{I} maps and the difference between them. The color scales are the same as Figure\,\ref{fig:jack8masks}.}
    \label{fig:jack4masks}
\end{figure*}

Figure\,\ref{fig:jack4masks} shows the results of the Jackknife Test in Stokes \textit{I} maps for the reduction method presented in this work. Figure\,\ref{fig:jack8masks} then shows the results of the Jackknife Test, but using the 8$\arcsec$ pixel auto-generated masks. The upper rows in Figures\,\ref{fig:jack8masks} and \ref{fig:jack4masks} are the same and shows the Stokes \textit{I} map from a standard 4$\arcsec$ reduction. All of the figures have the same color map scale and the difference map is the odd map subtracted from the even map. Stokes \textit{Q} and \textit{U} emission is very weak in starless cores and so little difference was seen between the two methods.

In Figure \ref{fig:jack4masks}, there is some difference seen between the ‘even’ and ‘odd’ maps with the normal 4$\arcsec$ reduction, but this same difference can be seen in the regridded 8$\arcsec$ reduction, just slightly blurred due to the larger pixel sizes. The difference is most likely due to the group selection and would change with different grouping. However, in Figure \ref{fig:jack8masks}, the difference in the auto-generated 8$\arcsec$ mask reduction is very different from the 4$\arcsec$ reduction. Additionally, the difference is seen in the areas of extended emission that appear in the normal 8$\arcsec$ reductions but not in the 4$\arcsec$ reduction (as traced by the contours). This difference is why we raise concerns with blindly increasing the value of the \textit{pixsize} parameter in the data reduction and potentially producing artificial structures. A different Jackknife Test grouping may yield a different residual map, or one that is not so severe. Future reduction tests can be conducted to determine if it is a selection effect or in fact growth of non-astronomical signal.

\end{document}